\documentclass[a4paper,11pt,oneside]{article}

\usepackage[english]{babel}
\usepackage[T1]{fontenc}
\usepackage[utf8]{inputenc}

\usepackage[pdftex,unicode]{hyperref}
\hypersetup{pdftitle=Analyzing and Forecasting Success in the Men's Ice Hockey World (Junior) Championships Using a Dynamic Ranking Model}
\hypersetup{pdfauthor=Vladimír Holý}

\usepackage{tikz}

\usepackage{xcolor}
\definecolor{mygold}{rgb}{1.0, 0.84, 0.0}
\definecolor{mysilver}{rgb}{0.75, 0.75, 0.75}
\definecolor{mybronze}{rgb}{0.8, 0.5, 0.2}
\definecolor{mygray}{rgb}{0.75, 0.75, 0.75}
\definecolor{myblue}{rgb}{0.18, 0.33, 0.69}
\definecolor{myred}{rgb}{0.98, 0.35, 0.27}
\hypersetup{colorlinks=true, linkcolor=myblue, anchorcolor=myblue, citecolor=myblue, filecolor=myblue, urlcolor=myblue}

\usepackage{tikz}
\usetikzlibrary{arrows.meta, positioning}

\usepackage[margin=60pt]{geometry}
\usepackage[authoryear]{natbib}

\usepackage{amsmath}
\usepackage{amssymb}
\usepackage{graphicx}
\usepackage[group-minimum-digits=3]{siunitx}
\usepackage{booktabs}
\usepackage{multirow}
\usepackage{tablefootnote}
\usepackage{float}
\usepackage{caption}

\usepackage[normalem]{ulem}

\usepackage{pifont}

\usepackage[super]{nth}
\newcommand{\rth}[1]{\ensuremath{{#1^\text{th}}}}

\newcommand{\goldmedal}[0]{
\begin{tikzpicture}
\fill[fill=mygold] (0,0) circle (2.5mm);
\node at (0,0) {1};
\end{tikzpicture}
}

\newcommand{\silvermedal}[0]{
\begin{tikzpicture}
\fill[fill=mysilver] (0,0) circle (2.5mm);
\node at (0,0) {2};
\end{tikzpicture}
}

\newcommand{\bronzemedal}[0]{
\begin{tikzpicture}
\fill[fill=mybronze] (0,0) circle (2.5mm);
\node at (0,0) {3};
\end{tikzpicture}
}

\begin{document}

\begin{center}
{\Large \bfseries Analyzing and Forecasting Success in the Men's Ice Hockey World (Junior) Championships Using a Dynamic Ranking Model}
\end{center}

\begin{center}
{\bfseries Vladimír Holý} \\
Prague University of Economics and Business \\
Winston Churchill Square 4, 130 67 Prague 3, Czechia \\
\href{mailto:vladimir.holy@vse.cz}{vladimir.holy@vse.cz} \\
\end{center}

\noindent
\textbf{Abstract:}
What factors contribute to the success of national teams in the Men's Ice Hockey World Championships and the Men's Ice Hockey World Junior Championships? This study examines whether hosting the tournament provides a home advantage; the influence of past tournament performances; the impact of players' physical characteristics such as height, weight, and age; and the value of experience from the World Championships compared to the NHL and other leagues. We employ a dynamic ranking model based on the Plackett--Luce distribution with time-varying strength parameters driven by the score. The results show that experience in the IIHF tournaments outweighs experience in the NHL. Furthermore, in junior championships, there is a significant home advantage, and shorter, heavier players tend to have an edge. In senior championships, future success is linked to past achievements in both junior and senior championships, with younger teams performing better. Finally, we conduct a forecasting analysis to predict the probabilities of winning the tournament, earning a medal, and advancing to the playoff phase.
\\

\\

\\

\section{Introduction}
\label{sec:intro}

The Men's Ice Hockey World Championships (WC) are annual international tournaments organized by the International Ice Hockey Federation (IIHF). They are among the most prestigious events in the sport of ice hockey and feature national teams from around the world competing for the title of world champion. The IIHF also organizes the Men's Ice Hockey World Junior Championships (WJC) for players under 20 years old. This paper focuses on a statistical analysis of the results from both of these tournaments.

In the field of sports statistics, ice hockey has been analyzed from various perspectives. The existing literature predominantly covers the National Hockey League (NHL), but other leagues are occasionally analyzed as well. For example, \cite{Leard2011} utilized a probit model to analyze NHL games and found evidence of home advantage and game-to-game momentum, but no link between winning fights and winning games. \cite{Doyle2012} further focused on the sources of home advantage in the NHL using a panel regression on season-level data and a probit model on game-level data. \cite{Guerette2021} modeled the number of awarded penalties and found that NHL referees favored the home team when spectators were present. \cite{Thrane2023} conducted a similar study for the Norwegian league and found that there is a home advantage in the presence of spectators. \cite{Marek2014} modeled and predicted the outcomes of games in the Czech league using the bivariate Poisson distribution with exponentially weighted observations. \cite{Whelan2021} studied and extended the Bradley--Terry model in the context of ice hockey. \cite{Buttrey2011} estimated the rates at which NHL teams score and concede goals. \cite{Gu2019} developed a machine learning system to predict the outcomes of NHL games. \cite{Schulte2017} utilized the Markov game framework to evaluate player actions, locations, and team performance in the NHL. \cite{MacDonald2011} developed an adjusted plus-minus statistic to quantify the overall contribution of NHL players to their teams. \cite{Tarter2009} evaluated the long-term value of elite junior ice hockey players to their future NHL teams. \cite{VonAllmen2015} examined wage premia for European players in the NHL.

These studies mostly focus on the outcomes of individual matches or specific partial statistics. The study of \cite{Holy2022f}, however, offers a notable exception by modeling the final standings, i.e., the complete ranking of all participating teams. Prior to the work of \cite{Holy2022f}, a review of the ranking literature by \cite{Yu2019} highlighted the lack of dynamics in existing ranking models. Addressing this gap, \cite{Holy2022f} proposed a score-driven model based on the Plackett--Luce distribution, offering a universally applicable time series approach for rankings. This model was applied to the results of the Men's Ice Hockey World Championships. However, the application was more illustrative than in-depth, and the approach was purely time-series-based and incorporated only a single predictor variable: a dummy variable indicating the host country, which was ultimately found to be insignificant in the analysis.

This paper aims to build upon and extend the study of \cite{Holy2022f} by incorporating a broader range of predictor variables. In addition to focusing on the World Championships, we also examine the World Junior Championships and explore the relation between the two. Our first objective is to evaluate the impact of factors such as the home advantage, results from related tournaments, players' physical attributes, and players' experience on the success of national teams. These variables are crucial as they reflect the multifaceted nature of performance in international ice hockey, where success depends not only on team strategies and individual skills but also on contextual and physical factors. By analyzing these elements, we contribute to more informed decision-making in national team management and player development strategies. The inclusion of these variables necessitates several modifications to the model proposed by \cite{Holy2022f}. Consequently, we introduce a novel ranking model that features separate regression and dynamic components and can handle non-participating teams. A scheme of the model structure is presented in Figure \ref{fig:scheme}. Our second objective is to assess the forecasting performance of our approach. Since our main focus is on the ranking, the model naturally provides predictions for various derived probabilities, such as the likelihood of winning the tournament, earning a medal, or advancing to the playoff phase.

Among the key questions we seek to answer are: Do national teams exhibit stable strength levels over the long term, or do their strengths evolve more rapidly? Does the hosting team enjoy a home advantage? Does success in the World Junior Championship translate into success in the next World Championship? Similarly, does success in the World Under-18 Championship translate into success in the next World Junior Championship? Do overall physical characteristics, such as average height and weight, significantly impact performance? Is it advantageous to have a younger team in the World Championships? Is experience from the World Championships more valuable than experience from the NHL? Are there notable differences in the factors influencing performance in the World Junior Championships compared to the World Championships? How accurately can the results of the tournaments be predicted?

This study is significant for both its methodological contributions and practical implications. By refining the dynamic ranking model of \cite{Holy2022f} and incorporating a diverse range of predictor variables, it advances the statistical modeling of sports tournaments, with applications extending beyond ice hockey, while addressing gaps in the current literature. Practically, the insights gained can help national hockey organizations better prepare for international competitions. Furthermore, understanding the dynamics between junior and senior tournaments offers valuable information for developing young players and strategically planning hockey programs.

The remainder of the paper is structured as follows: In Section \ref{sec:back}, we highlight the specifics of the WC and WJC and present the analyzed dataset. In Section \ref{sec:meth}, we propose our dynamic ranking model and describe the methods of its estimation. In Section \ref{sec:res}, we report the results of the analysis and discuss their implications. We conclude the paper in Section \ref{sec:con}.

\begin{figure}
\resizebox{\textwidth}{!}{
\begin{tikzpicture}[
    box/.style={text width=8.2cm, align=left, inner xsep=0.6cm, inner ysep=0.3cm, outer sep=0pt},
    arrow/.style={-Latex, thick, shorten >=0.4cm, shorten <=0.2cm}
]

    \node[box, fill=myblue!15, label={[rotate=90, text=myblue, anchor=south]left:Long-Term}, anchor=north west] (long) 
    {\textbf{Fixed Effects} \\ Capturing long-term intrinsic skill level, institutional support, national culture, and fan base};

    \node[box, fill=myblue!15, label={[rotate=90, text=myblue, anchor=south]left:Medium-Term}, anchor=north west] (medium) at ([xshift=0cm, yshift=-0.5cm] long.south west)  {\textbf{Dynamic Component} \\ Modeling medium-term deviation from the long-term mean using mean-reverting dynamics};
    \node[box, fill=myblue!15, label={[rotate=90, text=myblue, anchor=south]left:Short-Term}, anchor=north west] (short) at ([xshift=0cm, yshift=-0.5cm] medium.south west) {\textbf{Hosting the Tournament Variable} \\ Determining the potential home advantage \\ \vspace{0.3cm} \textbf{Past Results Variables} \\ Relating to success in past U18, WJC, and WC \\ \vspace{0.3cm} \textbf{Physical Attributes Variables} \\ Assessing the impact of players' height, weight, and age \\ \vspace{0.3cm} \textbf{Experience Variables} \\ Assessing the impact of players' experience in IIHF, NHL, and other tournaments};

    \node[box, fill=myred!15, right=2cm of medium, label={[rotate=270, text=myred, anchor=south]right:Unobserved}] (strengths) {\textbf{Team Strengths} \\ Parameters of the Plackett--Luce distribution};
    \node[box, fill=myred!15, right=2cm of short, label={[rotate=270, text=myred, anchor=south]right:Observed}] (standings) {\textbf{Final Standings} \\ Complete ranking of all participating teams};

    \draw[arrow] (long.east) -- (strengths.west);
    \draw[arrow] (medium.east) -- (strengths.west);
    \draw[arrow] (short.east) -- (strengths.west);

    \draw[arrow] (strengths.south) -- (standings.north);
\end{tikzpicture}
}
\caption{A scheme of the model structure.}
\label{fig:scheme}
\end{figure}


\section{Background}
\label{sec:back}

\subsection{World Championships and Other Tournaments}
\label{sec:backHockey}

The Men's Ice Hockey World Championships were first officially held during the 1920 Summer Olympics. In 1930, the first championship separate from the Olympics took place. These championships continued to be held annually, with the Olympic hockey tournaments serving as the World Championships for those years until 1968. Since 1972, the World Championships have been held independently, even in Olympic years. During the three Winter Olympics in the 80s, however, the World Championships were not held. The tournament was also not held from 1940 to 1946 due to World War II. Recently, the 2020 tournament was canceled due to the COVID-19 pandemic. The 2024 tournament marked the 87th edition.

Currently, the World Championships feature 16 teams. The preliminary round consists of two groups, with seedings determined by the IIHF World Ranking. Each group includes 8 teams competing in a round-robin format, with the top 4 teams from each group advancing to the playoff round. This round begins with four quarter-final matches, and the four winning teams progress to the semi-finals. The tournament concludes with the final match and the third-place game. The final standings represent the complete ranking of all 16 participating teams. The ranks of the top 4 teams are determined by the playoffs, while the ranks of the remaining teams are primarily determined by the points earned from the matches in the preliminary round and the quarter-finals. The final standings are used to update the IIHF World Ranking. In addition to this Top Division, the tournament features lower-tier Divisions I through IV. The bottom two teams from the Top Division are relegated to Division I for the following year, while the top two teams from Division I are promoted to the Top Division.

The World Championships, along with the Olympic ice hockey tournament (also governed by the IIHF), are considered the most prestigious international ice hockey tournaments. The NHL, on the other hand, is the premier professional ice hockey league in North America and features many of the world's best players. While the World Championships and the Olympic tournament highlight international competition among national teams, the NHL represents the highest level of club competition in the world. The World Championships are typically held in May, following the NHL regular season but during the NHL playoffs. Consequently, some top NHL players are unable to participate in the World Championships due to NHL commitments or injuries. Some characteristics of both tournaments are listed in Table \ref{tab:leagues}. The two tournaments also differ in the size and shape of the rinks, as well as in some specific rules.

The World Championships are dominated by the so-called Big Six teams: Canada and the United States from North America, and Czechia, Finland, Russia, and Sweden from Europe. The Big Six teams collectively won 85 gold medals, while the World Championships were won by another team only twice.  The Czechia team is the successor to the Czechoslovakia team, which disbanded in 1992. At the same time, the Slovakia team also emerged, starting international play in the lowest pool and advancing to the top pool by 1996. The Russia team is the successor to the Soviet Union team, which disbanded in 1991, and briefly competed as the Commonwealth of Independent States (CIS) team. Following the dissolution of the Soviet Union; Belarus, Kazakhstan, Latvia, and Ukraine, among others, established their own national teams. The modern Germany team emerged after the unification of the West and East German teams in 1991 and is considered the successor to West Germany. In this paper, we consider Czechoslovakia/Czechia, the Soviet Union/CIS/Russia, and West Germany/Germany as single teams, referring to them by their latest designation. Other teams regularly participating in the World Championships include Austria, Denmark, France, Italy, Norway, Slovenia, and Switzerland. Occasionally, teams from Great Britain, Hungary, Japan, and Poland also participate.

Each year, a different country hosts the World Championships. The most frequent host countries are Czechia and Sweden, each having hosted 11 times, followed by Finland and Switzerland, each having hosted 10 times. Occasionally, the hosting is shared between two or even three countries. The World Championships have been hosted outside Europe only four times: once in Canada and three times in the United States. It is evident that the tournament is popular in Europe, while North America prioritizes the NHL. This is especially true for the United States, which often uses the World Championships as an opportunity to give experience to younger players.

The IIHF also organizes tournaments for younger men. From 1974 to 1976, three unofficial junior tournaments were held for players under 20. Since 1977, the official Men's World Junior Championships, also known as the Men's World Under-20 Championships, has been held by the IIHF. The 2024 tournament marked the 48th edition. The WJC tournament usually takes place at the turn of December and January. The WJC tournament is especially popular in Canada, which has hosted it 17 times. In total, the tournament was held 22 times in North America and 26 times in Europe\footnote{Excluding the unofficial tournaments in 1974 in the Soviet Union, in 1975 in Canada and the United States, and in 1976 in Finland.}. For even younger players, there is the Men's World Under-18 Championships (U18). The tournament started in 1999, and the 2024 edition marked its 25th iteration. The U18 tournament usually takes place at April. The U18 tournament is dominated by the United States, which has won 11 out of 25 times. Both the WJC and U18 tournaments are organized according to a system similar to that of the World Championships. 

\begin{table}
\caption{An overview of selected major men's ice hockey tournaments and leagues. The statistics are reported for the 2023/2024 season (with the exception of the 2022 Olympics) and cover the number of participating teams, the average number of games per team, and the average attendance per game. The source of data is IIHF (\url{www.iihf.com}) and Alliance of European Hockey Clubs (\url{www.eurohockeyclubs.com}).}
\label{tab:leagues}
\centering
\resizebox{\textwidth}{!}{
\begin{tabular}{lllrrrr}
\toprule
\multicolumn{2}{l}{Tournament / League} & Country & Founded & Teams & Games & Attendance \\
\midrule
OG & Ice Hockey at Olympics & Worldwide & 1920 & 12 & 5.00 & 915\tablefootnote{The low attendance at the 2022 Winter Olympics in Beijing was caused by COVID-19 restrictions.} \\
\\ 
WC & World Championships & Worldwide & 1920 & 16 & 8.00 & 12\,464 \\
WJC & World Junior Championships & Worldwide & 1977 & 10 & 5.80 & 5\,885 \\
U18 & World Under-18 Championships & Worldwide & 1999 & 10 & 5.80 & 1\,826 \\
\\ 
NHL & National Hockey League & United States \& Canada & 1917 & 32 & 87.56 & 17\,452 \\
AHL & American Hockey League & United States \& Canada & 1936 & 32 & 77.13 & 5\,861 \\
\\ 
KHL & Kontinental Hockey League & Russia \& Other Countries & 2008 & 23 & 74.96 & 6\,623 \\
SHL & Swedish Hockey League & Sweden & 1975 & 14 & 58.71 & 6\,136 \\
SML & Liiga & Finland & 1975 & 15 & 66.00 & 4\,568 \\
NL & National League & Switzerland & 1938 & 14 & 58.71 & 7\,130 \\
DEL & Deutsche Eishockey Liga & Germany & 1994 & 14 & 57.71 & 7\,162 \\
ELH & Czech Extraliga & Czechia & 1993 & 14 & 59.71 & 5\,562 \\
\bottomrule
\end{tabular}
}
\end{table}

\subsection{Analyzed Data Sample}
\label{sec:backData}

We analyze the results of the World Championships from the 43rd edition in 1976 to the 87th edition in 2024. The tournament did not take place in 1980, 1984, 1988, and 2020, as discussed in Section \ref{sec:backHockey}. Therefore, we have 45 observations. The number of participating teams varied over the years: 8 teams from 1976 to 1991, 12 teams from 1992 to 1997, and 16 teams since 1998. A total of 27 teams have participated in the World Championships since 1976. However, the teams from the Netherlands, Romania, and South Korea each participated only once and finished last in their respective tournaments. We remove these teams from our sample, as they would be considered to have a strength of minus infinity. This issue is further discussed in Section \ref{sec:methLik}. Consequently, we retain 24 teams in our sample. Table \ref{tab:teams} provides a complete list of participating teams. The year 1976 was selected as the starting year for our data sample for several reasons: The 1976 tournament was the first to officially feature professional players\footnote{Prior to 1976, some nations, such as the Soviet Union, circumvented the rules by using professional players, whom they listed as military personnel.} and can thus be considered as the start of the modern era of World Championships. The tournament also underwent structural changes, with the number of participating teams increasing from 6 in 1975 to 8 in 1976. Additionally, the World Junior Championships began in the mid-1970s. The study by \cite{Holy2022f} analyzes the results of the World Championships from the 62nd edition in 1998 to the 83rd edition in 2019, with only 22 observations. Our analysis thus utilizes a much broader time frame, with double the number of observations.

Furthermore, we analyze the World Junior Championships from the 1st official edition in 1977 to the 48th edition in 2024. Since the tournament has taken place every year since its inception, we have a total of 48 observations\footnote{No Top Division tournaments were canceled due to COVID-19. However, all games in the 2021 and 2022 tournaments were held behind closed doors with no spectators. The 2022 tournament was also postponed from December/January to August.}. The number of participating teams was 8 from 1977 to 1995 and increased to 10 starting in 1996. A total of 19 teams have participated over the years. We exclude the teams from Austria, France, Japan, and Ukraine, as they form a separate group that never ranked higher than any of the remaining teams, which would result in a strength of minus infinity (see Section \ref{sec:methLik} for more details). Consequently, we retain 15 teams in our sample. Table \ref{tab:teams} provides a complete list of participating teams. The number of observations and time frame for our World Junior Championships sample roughly correspond to those of our World Championships sample, although the number of teams is much lower.

\begin{table}
\caption{The number of times a team has hosted, the number of times a team has participated, the mean reciprocal rank, the best rank, and the worst rank in the World Championships (43rd--87th editions) and the World Junior Championships (1st--48th editions). Only ranks from years when the team participated are included. Teams removed from the analysis are displayed in gray. The source of the data is IIHF (\url{www.iihf.com}).}
\label{tab:teams}
\centering
\begin{tabular}{lrrrrrcrrrrr}
\toprule
& \multicolumn{5}{c}{World Championships} & & \multicolumn{5}{c}{World Junior Championships} \\  \cmidrule(l{3pt}r{3pt}){2-6} \cmidrule(l{3pt}r{3pt}){8-12}
Team & Host & Part. & MRR & Best & Worst & & Host & Part. & MRR & Best & Worst \\ 
\midrule
Austria       &    4 &   22 & 0.083 &    8 &   16 & & \color{mygray} 0 & \color{mygray} 6 & \color{mygray} 0.106 & \color{mygray} 8 & \color{mygray} 10 \\ 
Belarus       &    1 &   20 & 0.099 &    6 &   15 & &    0 &    8 & 0.103 &    9 &   10 \\ 
Canada        &    1 &   44 & 0.453 &    1 &    8 & &   17 &   48 & 0.626 &    1 &    8 \\ 
Czechia       &    6 &   45 & 0.438 &    1 &    8 & &    6 &   48 & 0.286 &    1 &    7 \\ 
Denmark       &    1 &   21 & 0.089 &    8 &   14 & &    0 &    7 & 0.123 &    5 &   10 \\ 
East Germany  &    0 &    4 & 0.135 &    6 &    8 \\ 
Finland       &    8 &   45 & 0.334 &    1 &    8 & &    6 &   48 & 0.318 &    1 &    9 \\ 
France        &    1 &   25 & 0.086 &    8 &   16 & & \color{mygray} 0 & \color{mygray} 1 & \color{mygray} 0.100 & \color{mygray} 10 & \color{mygray} 10 \\ 
Germany       &    5 &   42 & 0.147 &    2 &   15 & &    2 &   32 & 0.133 &    5 &   10 \\ 
\\ 
Great Britain &    0 &    5 & 0.072 &   12 &   16 \\ 
Hungary       &    0 &    3 & 0.065 &   15 &   16 \\ 
Italy         &    1 &   23 & 0.094 &    6 &   16 \\ 
Japan         &    0 &    7 & 0.064 &   14 &   16 & & \color{mygray} 0 & \color{mygray} 1 & \color{mygray} 0.125 & \color{mygray} 8 & \color{mygray} 8 \\ 
Kazakhstan    &    0 &   12 & 0.074 &   10 &   16 & &    0 &    8 & 0.121 &    6 &   10 \\ 
Latvia        &    3 &   27 & 0.110 &    3 &   13 & &    0 &    9 & 0.115 &    7 &   10 \\ 
Netherlands   & \color{mygray} 0 & \color{mygray} 1 & \color{mygray} 0.125 & \color{mygray} 8 & \color{mygray} 8 \\ 
Norway        &    1 &   28 & 0.096 &    6 &   15 & &    0 &    9 & 0.121 &    6 &   10 \\ 
Poland        &    1 &    7 & 0.105 &    7 &   16 & &    0 &    6 & 0.124 &    7 &   10 \\ 
\\ 
Romania       & \color{mygray} 0 & \color{mygray} 1 & \color{mygray} 0.125 & \color{mygray} 8 & \color{mygray} 8 \\ 
Russia        &    5 &   42 & 0.504 &    1 &   11 & &    4 &   45 & 0.557 &    1 &    7 \\ 
Slovakia      &    2 &   28 & 0.190 &    1 &   14 & &    0 &   29 & 0.155 &    3 &    9 \\ 
Slovenia      &    0 &   10 & 0.067 &   13 &   16 \\ 
South Korea   & \color{mygray} 0 & \color{mygray} 1 & \color{mygray} 0.062 & \color{mygray} 16 & \color{mygray} 16 \\ 
Sweden        &    6 &   45 & 0.440 &    1 &    9 & &    7 &   48 & 0.327 &    1 &    8 \\ 
Switzerland   &    3 &   31 & 0.174 &    2 &   12 & &    1 &   37 & 0.151 &    3 &    9 \\
Ukraine       &    0 &    9 & 0.083 &    9 &   16 & & \color{mygray} 0 & \color{mygray} 4 & \color{mygray} 0.106 & \color{mygray} 8 & \color{mygray} 10 \\ 
United States &    0 &   44 & 0.199 &    3 &   13 & &    6 &   48 & 0.327 &    1 &    8 \\ 
\bottomrule
\end{tabular}
\end{table}

\subsection{Predictor Variables}
\label{sec:backVar}

For the analysis of both the World Championships and World Junior Championships, we utilize the following variables. As \cite{Holy2022f}, we define a dummy variable indicating whether the team is hosting the tournament. If multiple teams are hosting, each is assigned the value of 1, i.e., they are considered full-fledged hosts. Hosting teams could enjoy home crowd support, familiarity with the venue, and favorable scheduling. On the other hand, they might face increased pressure. The location of the tournament is also linked to the size and shape of the rinks, which differ between Europe and North America. While the WJC is occasionally played on both European and North American rinks, the WC is primarily held on European rinks. The source of the hosting data is the IIHF (\url{www.iihf.com}).

Next, we incorporate the results from the most recent World Championships, World Junior Championships, and World Under-18 Championships. Typically, the WJC is held first each season, followed by the U18, and finally the WC. In the analysis of the WC, we use the results from the WC of the previous year, along with the results from the current season's WJC and U18. For the WJC, we utilize the results from the WC, WJC, and U18 of the previous season. We define the three variables as the reciprocal ranks corresponding to the three respective tournaments. These values lie within the interval $(0,1]$. If a team did not participate in the tournament or if the tournament was not held, we assign a value of 0. We also include ranks from the unofficial WJC tournaments in 1974--1976 and from the lower divisions of the WC, WJC, and U18. All three tournaments feature national teams, and player pools can overlap across tournaments and years. Some dependency is thus expected; for example, success in the WJC could increase the chances of success in future WC. On the other hand, for example, it is unlikely that past WC results influence future WJC outcomes; nevertheless, we include this potential relation in our model and assess its statistical significance. The purpose of this is to use the same set of variables for both tournaments and to conduct a sanity check. The source of the tournament results data is IIHF (\url{www.iihf.com}).

Another set of variables captures the physical characteristics of players; specifically height, weight, and age. Height is recorded in centimeters, weight in kilograms, and age in years. Since we require variables that represent the entire team, we calculate the averages across all players. Admittedly, this approach results in some loss of detailed information, as goaltenders, defensemen, and forwards often necessitate different physical profiles. However, these averages provide a general sense of the team's physical makeup, which can still yield valuable insights. Height and weight are included because they may influence a player’s style of play and effectiveness in different situations. Shorter players often have a lower center of gravity, which can enhance their balance and ability to make quick turns and cuts on the ice. On the other hand, taller players generally have longer reach and their size can make them harder to move off the puck and more effective in body checks. Lighter players thrive in speed, using their finesse to evade defenders and create offensive opportunities. On the other hand, heavier players excel in physical play, using their size and strength to dominate in battles for the puck, deliver hard hits, and protect the net. Additionally, a higher weight relative to height might suggest a stronger or more muscular build, which is particularly relevant in high-stakes international tournaments. Age serves as a proxy for both physical and mental maturity. Younger players may bring agility and innovative strategies, while older players often provide experience and composure under pressure. The combination of these physical characteristics is likely to influence team dynamics and overall performance, making them crucial for understanding factors that drive success in international ice hockey. The source of the physical characteristics data is Elite Prospects (\url{www.eliteprospects.com}).

The final set of variables represents player experience. We define three variables based on the average number of games played in specific tournaments, calculated across all players. The first variable encompasses games in the World Championships and the Olympic ice hockey tournaments, which are both top international competitions sanctioned by the IIHF. The second variable covers games in the NHL, which is considered the premier national league globally. The third variable includes games played in other esteemed national leagues, namely the AHL, KHL, SHL, SML, NL, DEL, and ELH (as listed in Table \ref{tab:leagues}). Note that the World Championships and Olympic tournaments feature significantly fewer games compared to national leagues. For the average number of games per team in the 2023/2024 season, see Table \ref{tab:leagues}. Regarding the NHL, there is an issue where players from the top-performing NHL teams do not participate in the WC due to the overlap between the NHL playoffs and the WC. This may influence the interpretation of the NHL experience variable, as a player in the WC with significant NHL experience does not necessarily imply that the player has been particularly successful in the NHL. In general, studying player experience is crucial because it provides valuable insights into a player's ability to perform under various conditions, adapt to different styles of play, and contribute to team dynamics. Experience can be related to a player's consistency, decision-making, and mental resilience. More experienced players can have a deeper understanding of game strategy, better situational awareness, and the ability to handle high-pressure situations effectively. Additionally, experienced players can serve as mentors to younger or less experienced teammates, helping to improve overall team cohesion and leadership. The source of the experience data is Elite Prospects (\url{www.eliteprospects.com}).

\section{Methodology}
\label{sec:meth}

\subsection{Plackett--Luce Distribution}
\label{sec:methDistr}

For the underlying probability model for rankings, we use the Plackett--Luce distribution, introduced by \cite{Plackett1975} and \cite{Luce1959}. This model describes the probability of observing a specific ranking of a set of items, or teams in our context, based on their associated parameters. Suppose we have $N$ teams, and let $y = \left( y(1), \ldots, y(N) \right)$ be a permutation of $\{1,\ldots,N\}$ representing a specific ranking of these teams. Here, $y(i)$ indicates the rank assigned to team $i$. The inverse of a ranking, known as an ordering, is given by $y^{-1} = \left( y^{-1}(1), \ldots, y^{-1}(N) \right)$. Here, $y^{-1}(r)$, also denoted as $\rth{r}$, indicates the team with rank $r$.

Each team $i$ has an associated parameter $f_i$, which can be interpreted as the team's strength or skill. The probability mass function of the Plackett--Luce distribution is given by
\begin{equation}
\label{eq:probmass}
\mathrm{P} \left[ y \middle| f \right] = \prod_{r=1}^{N} \frac{\exp {f_\rth{r}}}{ \sum_{s=r}^n \exp {f_\rth{s}} }.
\end{equation}
For example, for $n=3$ teams, the probability probability mass function is simply
\begin{equation}
\label{eq:probmass3}
\mathrm{P} \left[ y \middle| f \right] = \frac{\exp f_{\nth{1}}}{\exp f_{\nth{1}} + \exp f_{\nth{2}} + \exp f_{\nth{3}}} \frac{\exp f_{\nth{2}}}{\exp f_{\nth{2}} + \exp f_{\nth{3}}} \frac{\exp f_{\nth{3}}}{\exp f_{\nth{3}}}.
\end{equation}
This formula captures the idea that the probability of each team being ranked in a particular position is proportional to its strength parameter relative to the teams that have not yet been ranked. In this parameterization, also used by \cite{Holy2022f}, the strength parameters $f_i$ are allowed to take any real value. Note that in this model, the probability mass function \eqref{eq:probmass} is invariant to the addition of a constant to all strength parameters $f_i$. We further discuss this issue in Section \ref{sec:methLik}.

The log-likelihood function for a single realization is given by
\begin{equation}
\label{eq:loglik}
\ell \left( f \middle| y \right) = \sum_{i=1}^{N} f_i - \sum_{r=1}^{N} \ln \left( \sum_{s=r}^N \exp f_{\rth{s}} \right).
\end{equation}
Continuing with our three-team example, we have
\begin{equation}
\label{eq:loglik3}
\ell \left( f \middle| y \right) = f_{\nth{1}} + f_{\nth{2}} - \ln \left( \exp f_{\nth{1}} + \exp f_{\nth{2}} + \exp f_{\nth{3}} \right) - \ln \left( \exp f_{\nth{2}} + \exp f_{\nth{3}} \right).
\end{equation}
The score function, i.e.\ the gradient of the log-likelihood function, is given by
\begin{equation}
\label{eq:score}
\nabla_i \left( f \middle| y \right) = 1 - \sum_{r=1}^{y(i)} \frac{\exp{f_i}}{\sum_{s=r}^N \exp{f_\rth{s}} }, \qquad i = 1, \ldots, N.
\end{equation}
For three teams, it is
\begin{equation}
\label{eq:score3}
\begin{aligned}
\nabla_{\nth{1}} \left( f \middle| y \right) &= 1 - \frac{\exp{f_{\nth{1}}}}{\exp{f_{\nth{1}}} + \exp{f_{\nth{2}}} + \exp{f_{\nth{3}}}}, \\
\nabla_{\nth{2}} \left( f \middle| y \right) &= 1 - \frac{\exp{f_{\nth{2}}}}{\exp{f_{\nth{1}}} + \exp{f_{\nth{2}}} + \exp{f_{\nth{3}}}} - \frac{\exp{f_{\nth{2}}}}{\exp{f_{\nth{2}}} + \exp{f_{\nth{3}}}}, \\
\nabla_{\nth{3}} \left( f \middle| y \right) &= - \frac{\exp{f_{\nth{3}}}}{\exp{f_{\nth{1}}} + \exp{f_{\nth{2}}} + \exp{f_{\nth{3}}}} - \frac{\exp{f_{\nth{3}}}}{\exp{f_{\nth{2}}} + \exp{f_{\nth{3}}}}. \\
\end{aligned}
\end{equation}
In general, the score has an expected value of zero, and its variance is known as the Fisher information. The score indicates the direction in which the fit of the distribution, defined by the strength parameters $f$, can be improved to better accommodate the observed ranking $y$. A positive score $\nabla_i$ suggests that increasing the strength of team $i$ would better match the observed ranking, while a negative score suggests that decreasing the strength would better align with the observed ranking. A score of zero indicates that the current strength of team $i$ provides the optimal fit. As such, the score can be useful in dynamic models for updating the strength parameters after a new ranking occurs.

\subsection{Score-Driven Dynamics}
\label{sec:methDyn}

The previous section deals with the case of a single ranking $y$. Now, we model multiple rankings $y_t$ evolving over time $t=1,\ldots,T$. As \cite{Holy2022f}, we capture time dependence in rankings by a score-driven model. The class of score-driven models, also known as generalized autoregressive score (GAS) models and dynamic conditional score (DCS) models, was proposed by \cite{Creal2013} and \cite{Harvey2013}. These models utilize the score to drive the dynamics of time-varying parameters for arbitrary probability distributions, allowing for the construction of time series models suitable for a wide range of data types. Besides \cite{Holy2022f}, score-driven models were also utilized in sports statistics for modeling tennis by \cite{Gorgi2019} and for modeling association football by \cite{Koopman2019} and \cite{Lasek2021}. However, while these studies focus on modeling the outcomes of individual matches, \cite{Holy2022f} and our study concentrate on the rankings of entire tournaments. In contrast to the ranking model of \cite{Holy2022f}, our proposed model has three key distinctions.

First, we assume that not every team compete in the tournament each year. The model of \cite{Holy2022f} considers all teams each year but utilizes partial rankings, in which participating teams are ranked and the remaining teams are placed below them but remain unranked. A major issue we face with this approach is the need to include predictor variables for teams that do not actually participate. Some of these predictor variables may not be available or may be irrelevant. For example, if a team is not participating in the Top Division, which we model, it may still participate in a lower division but with a worse lineup, as the competition in lower divisions does not attract the best players. Another issue is the disqualification of teams for reasons unrelated to their sporting ability. This is the case with the suspension of Russia and Belarus from 2022 onward due to the Russian invasion of Ukraine.

Second, as our data sample is much larger than that in \cite{Holy2022f}, we must address the issue of tournaments not being held in certain years. Such missing observations influence the dynamics of the model. We propose not ignoring these time gaps but rather updating the strength parameters by adjusting them closer to their long-term values. This is related to the first issue, as the absence of a tournament in a given year is equivalent to all teams not participating in that year.

Third, as our focus is on predictor variables, we propose a regression model with a separate dynamic component. Note that a similar model was employed by \cite{Holy2024a} in the context of assessing technical efficiency of decision-making units using data envelopment analysis. \cite{Holy2022f} utilize a time series model in which predictor variables and dynamics are intertwined, similarly to the autoregressive moving-average model with exogenous variables (ARMAX model). In the model of \cite{Holy2022f}, predictor variables influence future strength parameters through the autoregressive term, thus having a lasting impact. In contrast, the autoregressive term in our proposed model is separate from predictor variables. Although predictor variables still play a role in future strength parameters through the lagged score term, they can be interpreted with respect to the current strength only. As we utilize many predictor variables related to the lineup (e.g., average age, average experience), the proposed model structure is a natural choice.

Let us formally define the proposed dynamic ranking model. For each time $t = 1, \ldots, T$, we label the set of participating teams as $\mathcal{P}_t$. The number of participating teams can vary each year. The ranking of participating teams at time $t$ is denoted as $y_t$. We assume that $y_t$ follows the Plackett--Luce distribution with strength parameters $f_{i,t}$, $i \in \mathcal{P}_t$. Furthermore, we have $M$ predictor variables, and the $j$-th variable for team $i$ at time $t$ is denoted as $x_{i,t,j}$. For team $i$ participating at time $t$, $i \in \mathcal{P}_t $, we model its strength parameter as
\begin{equation}
\label{eq:gas}
f_{i, t} = \omega_i + \sum_{j=1}^M \beta_{j} x_{i,t,j} + u_{i, t}, \qquad u_{i,t} = \varphi u_{i,t-1} + \mathbb{I}_{ \{ i \in \mathcal{P}_{t-1} \} } \alpha \nabla_i \left( f_{t - 1} \middle| y_{t - 1} \right),
\end{equation}
where $\mathbb{I}_{ \{ \cdot \} }$ denotes the indicator function. For non-participating teams, we do not compute their strength $f_{i,t}$, but we do keep track of the dynamic component $u_{i,t}$.

The parameters of the model are the fixed effects for each team $\omega_i$, the regression coefficients $\beta_j$ common for all teams, and the autoregressive coefficient $\varphi$ with the score coefficient $\alpha$, also common for all teams. This model structure resembles a panel regression with dynamic errors. The fixed effects capture all factors that do not change over time in the sample, such as intrinsic skill level, institutional support, national culture, and fan base. Of course, in the truly long run, everything changes. After all, more than 150 years ago, there was no ice hockey. The regression term represents the effect of the predictor variables on the concurrent strength, as discussed above. The remaining variation in strength is captured by the dynamic component, which can range from temporary to persistent depending on the value of $\varphi$. If $\varphi \in (-1, 1)$, the dynamic component is a mean-reverting process, with a long-term value of zero. Furthermore, if a team did not participate at time $t-1$, the score term in \eqref{eq:gas} is zero, which is the expected value of the score, and $u_t$ thus moves closer to zero. If $\varphi=1$, the dynamic component is a non-stationary process resembling a random walk. Finally, the dynamic component must be initialized, and we set $u_{i, 0} = 0$ for all teams.

\subsection{Maximum Likelihood Estimation}
\label{sec:methLik}

The strength parameters are deterministic functions of the lagged ranking, the lagged strength parameters, and the contemporaneous predictor variables. Therefore, the model is observation-driven according to the classification of \cite{Cox1981}. It can be straightforwardly estimated using the maximum likelihood method. Specifically, the estimates of all coefficients $\theta = (\omega_1, \ldots, \omega_N, \beta_1, \ldots, \beta_M, \varphi, \alpha)^\intercal$ can be obtained by maximizing the log-likelihood function
\begin{equation}
\mathcal{L}(\theta) = \sum_{t=1}^T \ell \left( f_t \middle| y_t \right),
\end{equation}
where $\ell(\cdot | \cdot)$ is given by \eqref{eq:loglik} and $f_t = (f_{1,t}, \ldots, f_{N,t})^\intercal$ is given by \eqref{eq:gas}. This is a nonlinear optimization problem, and any general-purpose algorithm can be used to solve it. In order for $\mathcal{L}$ to have a unique global maximum, two issues have to be addressed.

First, the fixed effects $\omega_1, \ldots, \omega_N$ are not identified as $\ell(f_t | y_t)$ is invariant to the addition of a constant to all $\omega_1, \ldots, \omega_N$. Therefore, we employ the standardization
\begin{equation}
\label{eq:standard}
\sum_{i=1}^N \omega_i = 0, \qquad \text{or equivalently,} \qquad \omega_N = - \sum_{i=1}^{N-1} \omega_i.
\end{equation}
We can thus exclude $\omega_N$ from the estimation and then derive it from \eqref{eq:standard}, without loss of generality. Note that if no predictor variables were included, this would also implicate that $\sum_{i=1}^N f_{i,t} = 0$ for all $t$ (assuming $u_{i,0}=0$ for all $i$).

Second, as stated by \cite{Hunter2004}, it is necessary that for any possible partition of teams into two non-empty subsets, at least one team in the second subset is ranked higher than at least one team in the first subset. If this condition does not hold, there would be a non-empty set of teams that are always better than the remaining teams. Conversely, there would also be a non-empty set of teams that are always worse than the remaining teams. The fixed effects of the `best' teams would then tend to $\infty$, while the fixed effects of the `worst' teams would tend to $-\infty$. Although \cite{Hunter2004} formulated this condition for the static case, it applies in the same way to our dynamic case. It is easy to verify this condition in practice. If violated, the analysis must either be performed on a subset of teams for which this condition holds, or the optimization problem must be suitably reformulated, e.g., as further discussed in Section \ref{sec:methPen}.

If the fixed effects are standardized and the partition condition of \cite{Hunter2004} holds, there are additional conditions required for the estimator to be consistent and asymptotically normal. Besides several regularity assumptions, the key condition is filter invertibility, which, among other things, means that the effect of past strength parameters diminishes over time. Specifically, it is necessary that $\varphi \in (-1,1)$. However, the derivation of all conditions ensuring filter invertibility, as well as further conditions for consistency and asymptotic normality, is quite complex and beyond the scope of this paper. For the general asymptotic theory regarding score-driven models, we refer to \cite{Blasques2014a}, \cite{Blasques2018}, and \cite{Blasques2022}.

The standard errors of the estimated coefficients $\hat{\theta}$ can be estimated using the empirical Hesssian of the log-likelihood as
\begin{equation}
\hat{\mathrm{se}} \left[ \hat{\theta} \right] = \sqrt{ \mathrm{diag} \left[ - T \cdot H \left(\hat{\theta} \right)^{-1} \right] }, \qquad \text{where} \qquad H(\theta) = \frac{\partial^2 \mathcal{L}(\theta)}{\partial \theta \partial \theta^\intercal}.
\end{equation}
The confidence intervals and significance (p-values) of the estimated coefficients can then be obtained by approximating the unknown finite-sample distribution with the asymptotic distribution, which is the normal distribution.

\subsection{Penalized Estimation}
\label{sec:methPen}

When the goal of the analysis is forecasting, it may be appropriate to include a regularization term and maximize a penalized version of the log-likelihood. Using regularization results in biased estimates; however, it can improve predictive accuracy by preventing overfitting. Note that because the estimates are biased, standard errors are not particularly informative in this case. In our model, regularization also removes the need for the partition condition of \cite{Hunter2004}, as the optimal fixed effects are always finite when the regularization term is present. In the context of sports statistics, \cite{Groll2015} use L$_1$ (Lasso) penalization, while \cite{Lasek2021} use L$_2$ (ridge) penalization. L$_1$ penalization tends to produce sparse models by driving some coefficients exactly to zero, while L$_2$ penalization tends to shrink all coefficients towards zero but not exactly to zero, often resulting in more stable models.

For our purposes, L$_2$ penalization is more suitable. Let $\lambda \geq 0$ denote the regularization parameter that controls the strength of the penalization. The log-likelihood function with L$_2$ penalization is then given by
\begin{equation}
\mathcal{L}_{\mathrm{Pen}}(\theta) = \sum_{t=1}^T \ell \left( f_t \middle| y_t \right) - \lambda \sum_{t=1}^T \sum_{i \in \mathcal{P}_t} f_{i, t}^2.
\end{equation}
An appropriate value of $\lambda$ can be selected using a grid search with rolling cross-validation.

\subsection{Computation and Software}
\label{sec:methComp}

The analysis was conducted using the statistical software R. The associated data and code are available at \url{github.com/vladimirholy/ice-hockey-wc}. The code is based on the gasmodel package of \cite{Holy2024b}, which facilitates the estimation, forecasting, and simulation of a variety of score-driven models, including the dynamic ranking model employing the Plackett--Luce distribution. This package was utilized in the analysis presented in \cite{Holy2022f}. However, our model introduces several modifications, as detailed in Section \ref{sec:methDyn}. While the package supports both joint and separate modeling of regression and dynamic components, it does not accommodate varying sets of participating teams over time. Additionally, it handles missing values, such as those arising from canceled tournaments, differently. To address these limitations, the provided code includes several simplified functions adapted from the original package. Optimization is carried out using the 
Sbplx algorithm from the nloptr package of \cite{Ypma2020}, a re-implementation of the Subplex algorithm of \cite{Rowan1990}.

\subsection{Limitations and Potential Extensions}
\label{sec:methLim}

There are several limitations in our methodology. The Plackett--Luce distribution is based on the assumption of the independence of irrelevant alternatives, which is related to the Luce's choice axiom. In our context, this means that the (internal) ranking of a set of teams is not influenced by the presence or absence of other teams. This may not always hold in practice, as there could be dependencies among teams; that is, some teams may consistently perform better or worse against others, which may not be fully explained by the fixed effects and predictor variables. However, the Plackett--Luce model can still serve as a useful tool due to its relative simplicity. More complex approaches that introduce a dependency structure among all teams could potentially be adopted; however, they would require a relatively small number of teams and a large number of rankings. See, e.g., \cite{Yu2019} for an overview of relevant ranking literature.

Another issue is the aforementioned partition condition of \cite{Hunter2004}. Although the two presented approaches address the problem of a nonexisting maximum of the likelihood function, neither approach is ideal: removing certain teams from the analysis involves artificial data omission, while using the penalized maximum likelihood method introduces bias into the coefficient estimates. The issue could also be approached from a Bayesian perspective. Using an appropriate prior, potentially incorporating additional data sources, could prove useful (not only) in situations where the condition of \cite{Hunter2004} is violated. We leave this topic for future research.

Finally, the proposed model could also be extended to account for ties. In WC and WJC tournaments, there is a system of tiebreakers (such as goal differences and head-to-head results) that determines the final standings, which do not contain ties. Thus, ties are not considered in our study. However, in applications of our model to other tournaments and sports, ties may occur. In that case, the Plackett--Luce distribution would need to be extended to handle tied rankings. See, e.g., the approach to ties in \cite{Firth2019}.

\section{Results}
\label{sec:res}

\subsection{Exploratory Data Analysis}
\label{sec:resExpl}

\begin{figure}
\centering
\includegraphics[width=\textwidth]{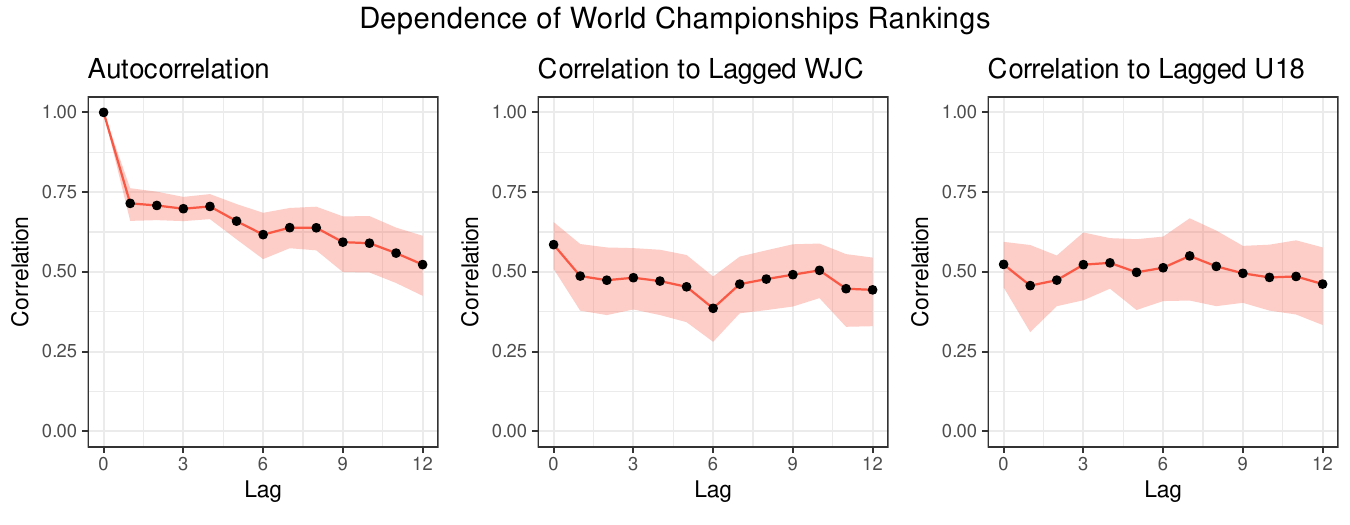}
\caption{The autocorrelation of World Championships ranks, the correlation of World Championships ranks with lagged World Junior Championships ranks, and the correlation of World Championships ranks with lagged World Under-18 Championships ranks. The plots include 95 percent confidence bands obtained using the bootstrapping method.}
\label{fig:autoWC}
\end{figure}

\begin{figure}
\includegraphics[width=\textwidth]{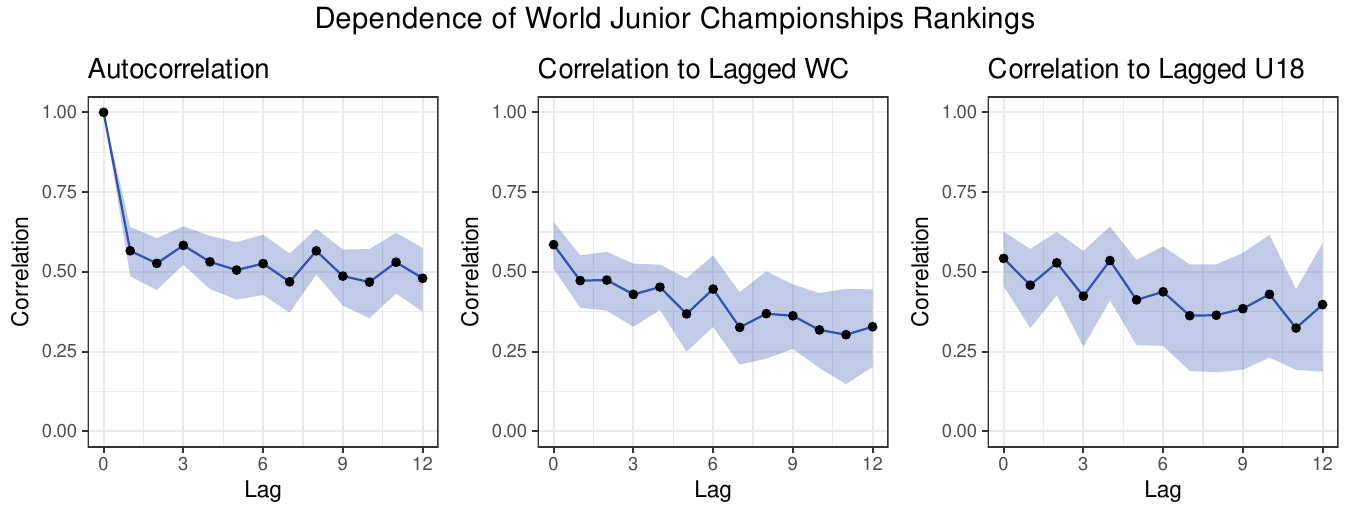}
\caption{The autocorrelation of World Junior Championships ranks, the correlation of World Junior Championships ranks with lagged World Championships ranks, and the correlation of World Junior Championships ranks with lagged World Under-18 Championships ranks. The plots include 95 percent confidence bands obtained using the bootstrapping method.}
\label{fig:autoU20}
\end{figure}

Figures \ref{fig:autoWC} and \ref{fig:autoU20} show the autocorrelations and cross-correlations of the World Championships and World Junior Championships rankings. For the WC, there is a strong positive autocorrelation, starting at 0.715 at lag 1 and gradually decreasing. For the WJC, the autocorrelation is lower, starting at 0.566 at lag 1. This suggests that teams in the WC exhibit more stable performance over time, while performance in the WJC tends to fluctuate. The concurrent correlation between WC and WJC rankings is 0.585, between WC and U18 rankings is 0.523, and between WJC and U18 rankings is 0.542. These correlations clearly indicate a connection between the three tournaments. Interestingly, there are also smaller positive correlations (below 0.5) between WJC and lagged WC. While this may not indicate direct causation, it suggests that the two tournaments could be linked through factors such as the overall quality and consistency of a country's hockey development system. These factors can be difficult to quantify (i.e., to use as predictor variables) but can be captured by the intercept together with the dynamic component in our model.

\newpage

\begin{figure}
\centering
\includegraphics[width=\textwidth]{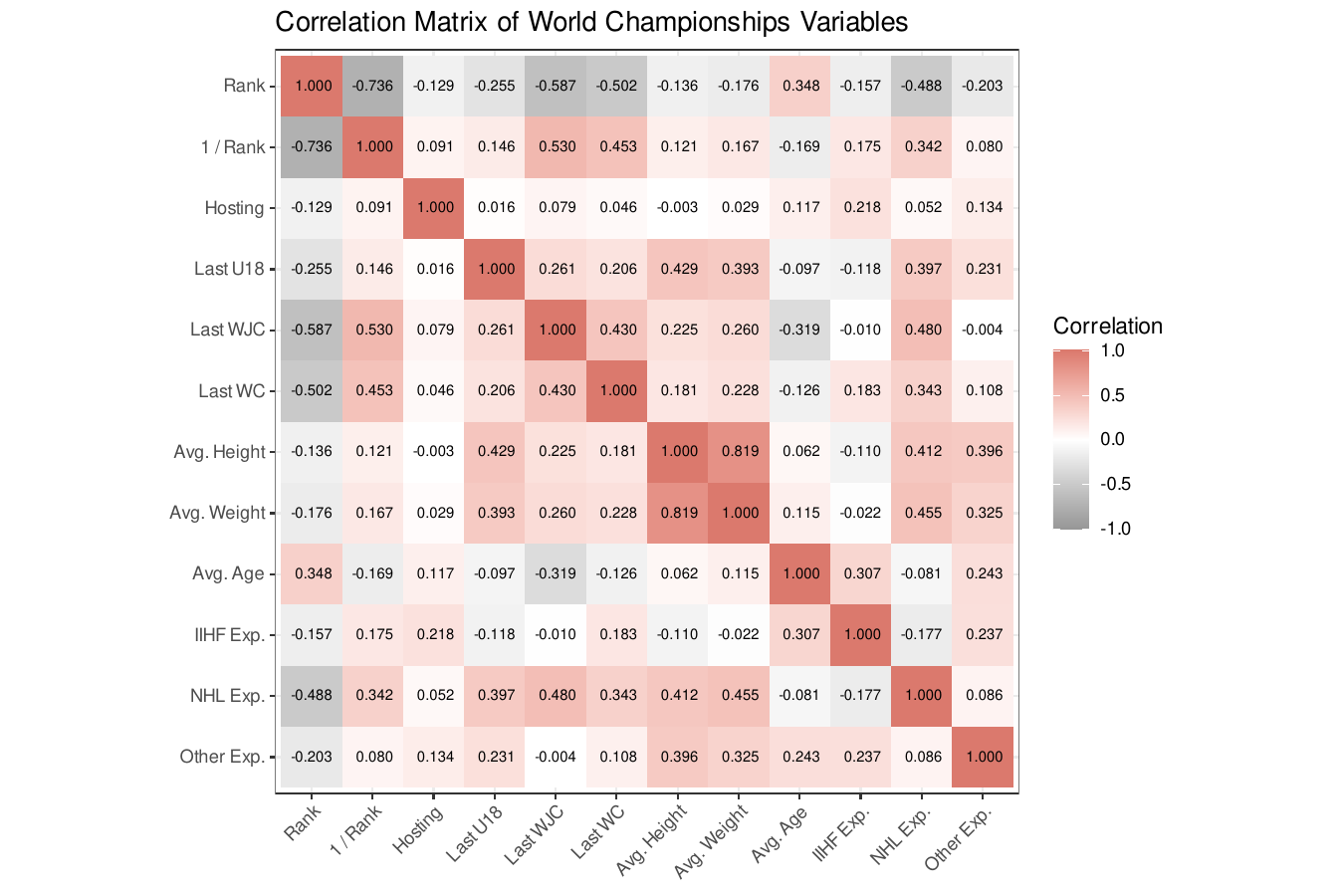}
\caption{The correlations between the rank, its reciprocal, and the predictor variables for the World Championships.}
\label{fig:corWC}
\end{figure}

\begin{figure}
\centering
\includegraphics[width=\textwidth]{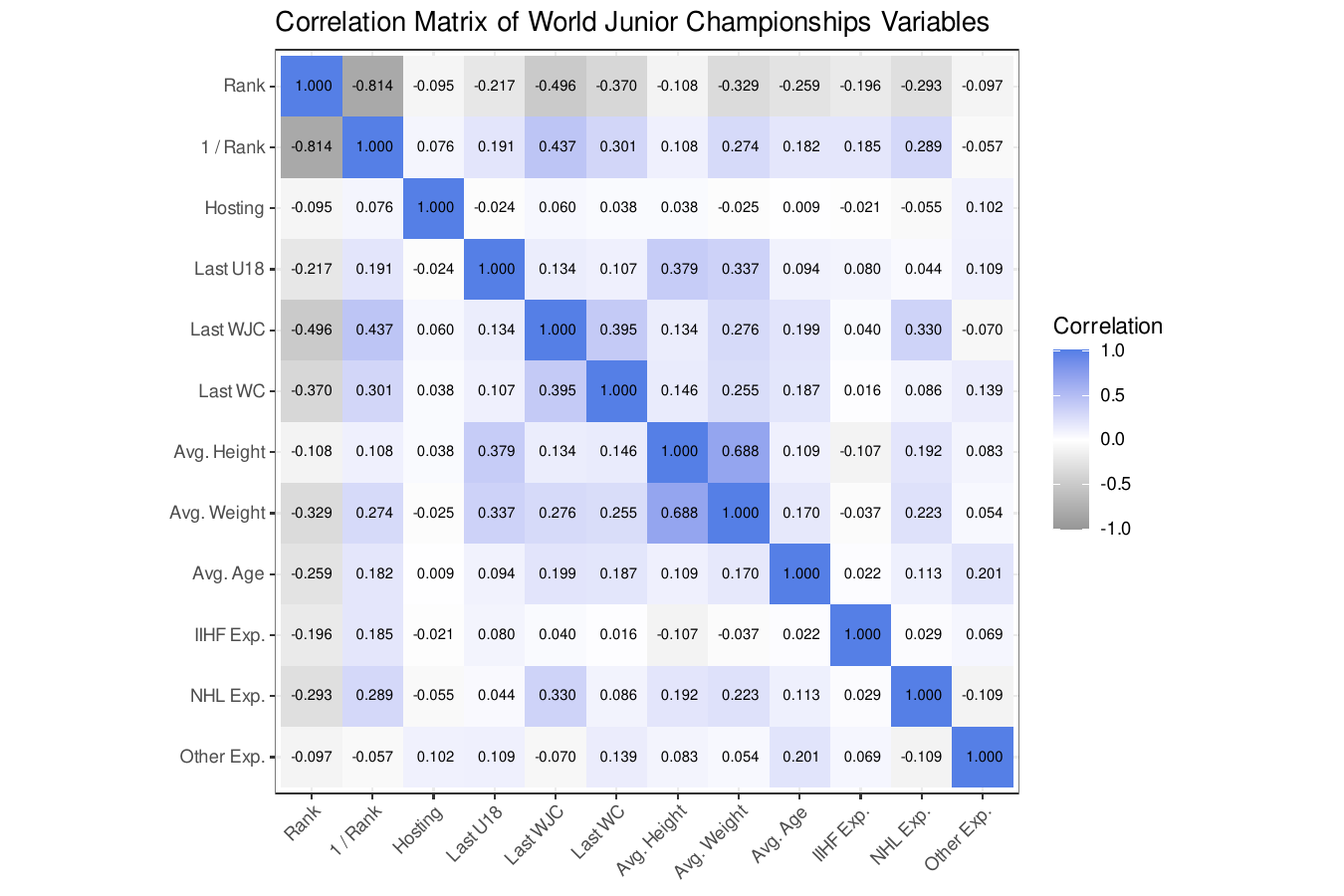}
\caption{The correlations between the rank, its reciprocal, and the predictor variables for the World Junior Championships.}
\label{fig:corU20}
\end{figure}

Figures \ref{fig:corWC} and \ref{fig:corU20} illustrate the correlations between the World Championships and World Junior Championships rankings and predictor variables. There is a modest negative correlation between ranking and hosting the tournament, with values of -0.129 for the WC and -0.095 for the WJC, suggesting a potential home advantage. In the WC, hosting is also positively correlated with IIHF experience (the WC and Olympic tournaments). Past success in the WC, WJC, and U18 tournaments is positively correlated with present success in both the WC and WJC (note that variables Last U18, Last WJC, and Last WC are in the form of reciprocal ranks). Successful teams tend to have higher average height and weight in both the WC and WJC. Height and weight are positively correlated with NHL experience but not with IIHF experience in both tournaments. They are also strongly correlated with each other. In the WC, successful teams tend to be younger on average, whereas in the WJC, where the age limit is 20, successful teams tend to be older on average, i.e, closer to that upper age limit. Age is positively correlated with IIHF experience but not with NHL experience in the WC. Player experience in any of the studied tournaments is positively correlated with team success. Notably, NHL experience has a much stronger correlation with rank compared to IIHF experience, with values of -0.488 versus -0.157 in the WC, and -0.293 versus -0.196 in the WJC. However, it should be noted that the vast majority of players in the WJC naturally have no IIHF or NHL experience. On average, teams in the WJC have 0.25 IIHF (WC or Olympics) games per player and 0.53 NHL games per player. More than half of the teams have no players with any IIHF experience, and the same applies to NHL experience. Consequently, variation in these two experience variables for the WJC is small. Finally, there is a small negative correlation of -0.177 between IIHF and NHL experience in the WC. Of course, the implications of this correlation analysis are quite limited, and the joint effect of all predictor variables should be further studied, as we do in the following section.

\subsection{Estimated Models}
\label{sec:resEst}

We turn to our proposed dynamic ranking model. We estimate seven models with various sets of predictor variables: the `static' model utilizes constant strength parameters; the `dynamic' model includes a dynamic component in the strength parameters but no predictor variables; the `tournament' model features the hosting variable and variables capturing the most recent results from the WC, WJC, and U18 tournaments; the `physical' model includes height, weight, and age variables along with the hosting variable; the `experience' model incorporates the three variables capturing player experience and the hosting variable; the `full' model includes all 10 predictor variables; and the `final' model utilizes the set of variables that leads to the lowest Akaike information criterion. For estimation, we use the maximum likelihood method with bounds $\varphi \in [0, 1)$ and $\alpha \geq 0$. We further discuss this issue and estimate unbounded models in Section \ref{sec:resOpti}.

\begin{figure}
\centering
\includegraphics[width=\textwidth]{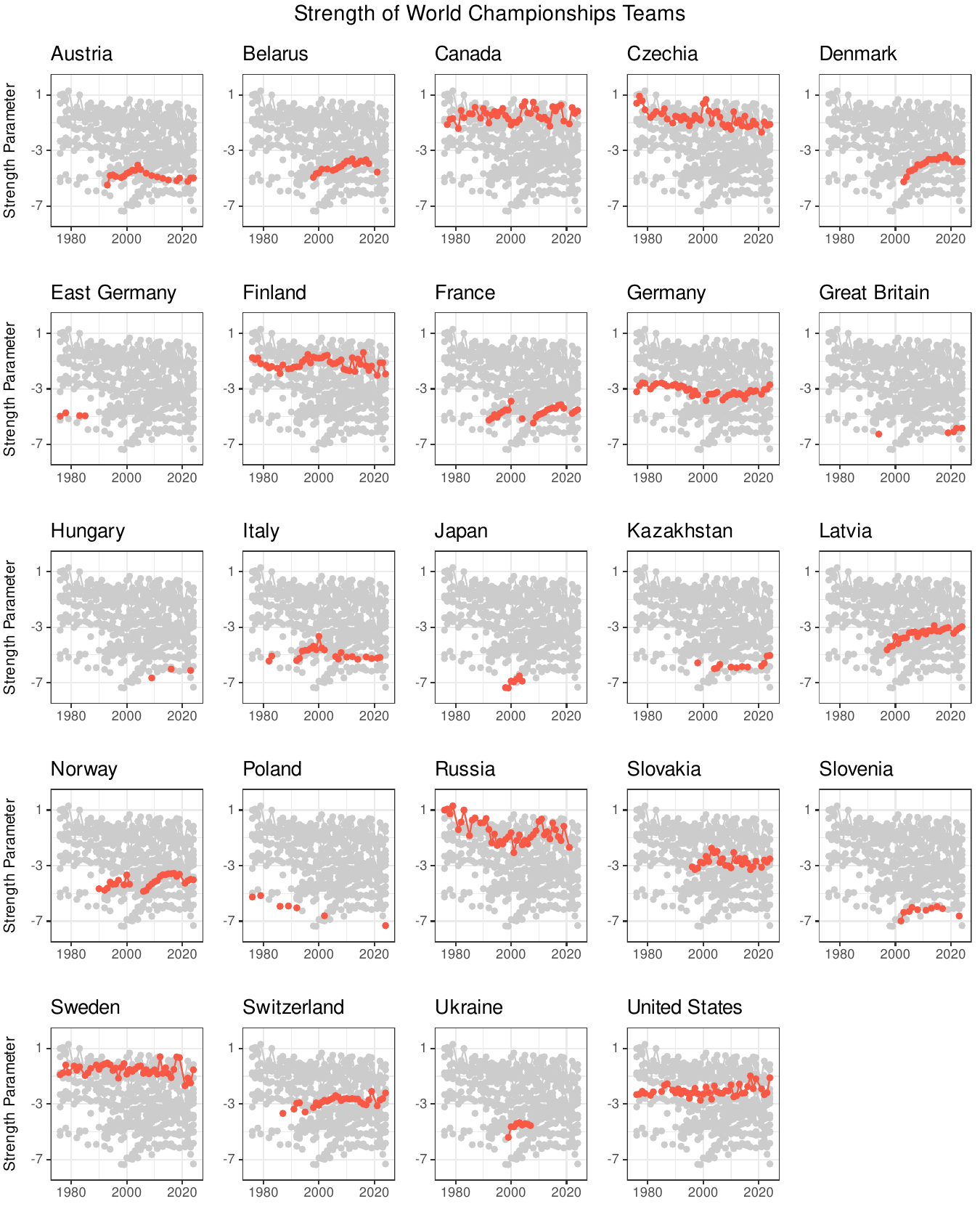}
\caption{The time-varying strength parameters for individual teams in the World Championships.  The strength parameters are obtained from the final model estimated by the maximum likelihood method. Only strength parameters for the years in which teams participated are included.}
\label{fig:strengthWC}
\end{figure}

First, we focus on analyzing the World Championships. The main results are reported in Table \ref{tab:estWC}, and the time-varying strength parameters are illustrated in Figure \ref{fig:strengthWC}. Based on the log-likelihood and Akaike information criterion, adding the dynamic component to the static model significantly improves the fit. The `dynamic' model is mean-reverting, with a value of 0.736 for the autoregressive coefficient, indicating medium persistence. In models with predictor variables, the hosting variable is positive, suggesting a slight increase in the strength of the hosting team, but it is statistically insignificant except for the `physical' model. Therefore, there is not enough evidence to support a distinct home advantage. Among the results from past tournaments, the reciprocal rankings from the most recent WJC and WC have a significantly positive effect, while the most recent U18 ranking is insignificant. Note that there is naturally less overlap of players between the WC and U18 than between the WC and WJC, and the U18 may also be viewed by national federations as part of an evaluation process rather than a culmination point. Adding the tournament variables reduces the value of the score coefficient, as they introduce similar information to the model. Regarding the physical attributes of players, the findings suggest that, on average, it is better to have shorter, heavier, and younger players. However, the effects of height and weight are not statistically significant in the `full' model. Age becomes statistically significant only when the experience variables are included, implying that younger players are preferable, provided they have the necessary experience. Among the experience variables, experience from IIHF tournaments is found to be much more important than NHL experience based on p-values, while experience from other national leagues is insignificant. The estimated coefficients for the experience variables should be interpreted in the context of the number of games played in IIHF tournaments versus the NHL, as reported in Table \ref{tab:leagues}. In a single NHL season, teams play, on average, 11 times more games than in a single WC tournament. However, the estimated coefficient for IIHF experience is 25 times larger than that for NHL experience, indicating that participation in a single WC tournament is much more valuable than participation in a single NHL season. The results regarding the NHL need to be interpreted with caution. As mentioned in Section \ref{sec:backHockey}, the WC typically takes place during the NHL playoffs. This means that players from the best-performing NHL teams do not participate in the WC. The NHL experience variable does not distinguish between regular-season and playoff games, nor does it capture success in games, only participation (as is the case with our other experience variables). Although, the best NHL teams play more games as they progress through the playoffs, and the number of games played can indirectly reflect success to a certain extent. Nevertheless, this may introduce some bias, as it is possible that the NHL experience variable is more closely related to less successful players, whereas the IIHF experience variable encompasses all players. Adding the experience variables reduces the value of the autoregressive coefficient, as they account for a portion of the observed autocorrelation. Despite this, we retain the dynamic component in the model. The experience variables improve the fit the most compared to the tournament and physical variables. The final model that minimizes the Akaike information criterion, and has all predictor variables significant at the 0.1 level, includes the reciprocal rank from the last WJC and WC tournaments, the average age, and the average IIHF and NHL experience.

\begin{figure}
\centering
\includegraphics[width=\textwidth]{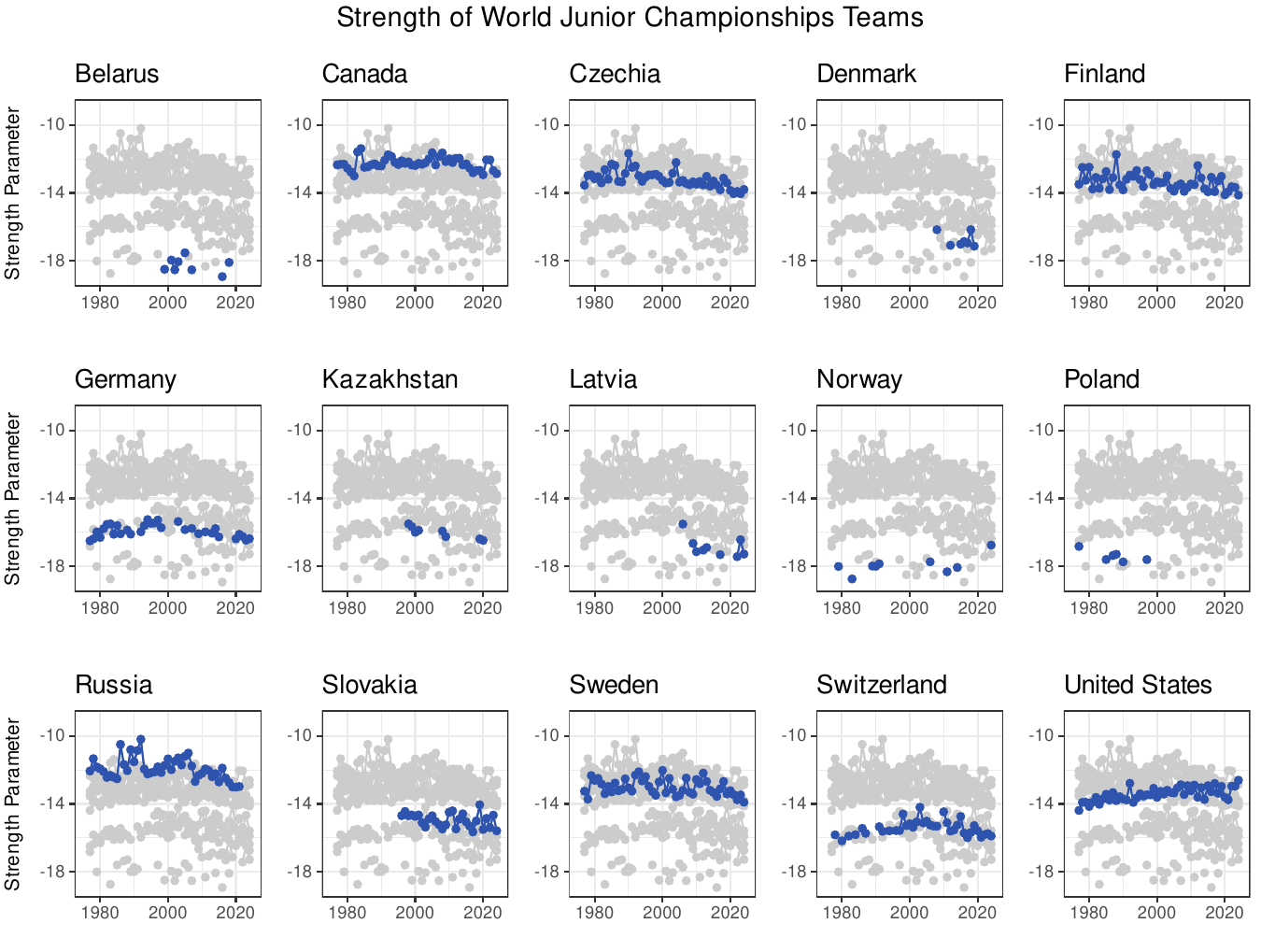}
\caption{The time-varying strength parameters for individual teams in the World Junior Championships. The strength parameters are obtained from the final model estimated by the maximum likelihood method. Only strength parameters for the years in which teams participated are included.}
\label{fig:strengthU20}
\end{figure}

Next, we focus on analyzing the World Junior Championships. The main results are reported in Table \ref{tab:estU20}, and the time-varying strength parameters are illustrated in Figure \ref{fig:strengthU20}. Similar to the WC analysis, adding the dynamic component to the static model significantly improves the fit. The hosting variable is consistently significant across all models, indicating a clear home advantage for the hosting team. This suggests that younger players participating in the WJC are more negatively affected by playing in a foreign environment compared to the more seasoned players in the WC. Note that although Canada is the most frequent host country, having hosted the WJC 17 times, there is sufficient variability in hosting locations, as the WJC was hosted 31 times outside of Canada. The results from the most recent WC, WJC, and U18 tournaments are all found to be insignificant. The lack of dependence on the most recent WC is expected, as the overlap of players between the last WC and the current WJC is minimal and quite rare. A similar situation occurs with the last WJC and the current WJC, as rosters often change dramatically from year to year. Some dependence on the last U18 tournament might be anticipated, given that many players transition from U18 to WJC, but it is ultimately found to be insignificant, even though the associated coefficient is the highest among the three tournaments. As in the case of the lack of dependency between WC and U18, it can be argued that U18 focuses on showcasing young talent and emphasizes player development, while WJC is often considered the highlight of junior careers. The average height and weight of players are significant, with the same signs as in the WC analysis: height has a negative impact, while weight has a positive impact (i.e., shorter and heavier players tend to perform better). Compared to the WC, these two variables have greater significance. This suggests that physical differences can play a major role early on but can be mitigated by continuous skill development. Age is found to be insignificant, likely due to the small variation in the average age level within the WJC. Among the experience variables, only experience from IIHF tournaments is found to be relevant, while experience from the NHL and other leagues is insignificant. As in the WC analysis, including experience improves the model fit the most. The final model is obtained by minimizing the Akaike information criterion, and, as in the WC analysis, all predictor variables are significant at the 0.1 level. The variables included in the final model are the hosting variable, average height, average weight, and average IIHF experience.

\begin{table}
\caption{The estimated coefficients with the standard errors in parentheses, along with the fitted log-likelihood and the Akaike information criterion for various models of the World Championships. The models are estimated by the maximum likelihood method. The fixed effects for individual teams are omitted. The best attained values of the log-likelihood and the Akaike information criterion are highlighted.}
\label{tab:estWC}
\centering
\begin{tabular}{lccccccc}
\toprule
 & Static & Dynamic & Tourn. & Phys. & Exp. & Full & Final  \\
\midrule
Hosting &  &  & 0.308 & 0.315$^{*}$ & 0.088 & 0.040 &  \\ 
   &  &  & (0.194) & (0.191) & (0.208) & (0.213) &  \\ 
   &  &  &  &  &  &  &  \\ 
Last U18 &  &  & -0.001 &  &  & 0.036 &  \\ 
   &  &  & (0.334) &  &  & (0.343) &  \\ 
   &  &  &  &  &  &  &  \\ 
Last WJC &  &  & 0.958$^{***}$ &  &  & 0.860$^{***}$ & 0.853$^{***}$ \\ 
   &  &  & (0.288) &  &  & (0.301) & (0.295) \\ 
   &  &  &  &  &  &  &  \\ 
Last WC &  &  & 0.853$^{***}$ &  &  & 0.699$^{**}$ & 0.674$^{**}$ \\ 
   &  &  & (0.308) &  &  & (0.315) & (0.313) \\ 
   &  &  &  &  &  &  &  \\ 
Avg. Height &  &  &  & -0.107 &  & -0.092 &  \\ 
   &  &  &  & (0.075) &  & (0.075) &  \\ 
   &  &  &  &  &  &  &  \\ 
Avg. Weight &  &  &  & 0.136$^{**}$ &  & 0.071 &  \\ 
   &  &  &  & (0.058) &  & (0.058) &  \\ 
   &  &  &  &  &  &  &  \\ 
Avg. Age &  &  &  & -0.070 &  & -0.222$^{***}$ & -0.191$^{***}$ \\ 
   &  &  &  & (0.055) &  & (0.070) & (0.065) \\ 
   &  &  &  &  &  &  &  \\ 
IIHF Exp. &  &  &  &  & 0.045$^{***}$ & 0.052$^{***}$ & 0.053$^{***}$ \\ 
   &  &  &  &  & (0.010) & (0.011) & (0.011) \\ 
   &  &  &  &  &  &  &  \\ 
NHL Exp. &  &  &  &  & -0.000 & 0.002 & 0.002$^{*}$ \\ 
   &  &  &  &  & (0.001) & (0.001) & (0.001) \\ 
   &  &  &  &  &  &  &  \\ 
Other Exp. &  &  &  &  & -0.001 & 0.001 &  \\ 
   &  &  &  &  & (0.001) & (0.001) &  \\ 
   &  &  &  &  &  &  &  \\ 
Autoreg. Coef. &  & 0.736$^{***}$ & 0.747$^{***}$ & 0.756$^{***}$ & 0.591$^{**}$ & 0.677$^{**}$ & 0.674$^{**}$ \\ 
   &  & (0.114) & (0.146) & (0.103) & (0.292) & (0.323) & (0.321) \\ 
   &  &  &  &  &  &  &  \\ 
Score Coef. &  & 0.186$^{***}$ & 0.078 & 0.190$^{***}$ & 0.136$^{**}$ & 0.039 & 0.040 \\ 
   &  & (0.060) & (0.063) & (0.061) & (0.069) & (0.070) & (0.068) \\ 
\midrule
Log-Lik. & -765.832 & -759.578 & -749.232 & -754.690 & -747.819 & \textbf{-734.524} & -735.989 \\ 
AIC & 1577.664 & 1569.155 & 1556.464 & 1567.380 & 1553.638 & 1539.047 & \textbf{1531.978} \\ 
\bottomrule
\midrule
\multicolumn{8}{r}{$^{***}p < 0.01$; $^{**}p < 0.05$; $^{*}p < 0.1$} \\
\end{tabular}
\end{table}

\begin{table}
\caption{The estimated coefficients with the standard errors in parentheses, along with the fitted log-likelihood and the Akaike information criterion for various models of the World Junior Championships. The models are estimated by the maximum likelihood method. The fixed effects for individual teams are omitted. The best attained values of the log-likelihood and the Akaike information criterion are highlighted.}
\label{tab:estU20}
\centering
\begin{tabular}{lccccccc}
\toprule
 & Static & Dynamic & Tourn. & Phys. & Exp. & Full & Final  \\
\midrule
  Hosting &  &  & 0.488$^{**}$ & 0.557$^{**}$ & 0.587$^{**}$ & 0.673$^{***}$ & 0.653$^{***}$ \\ 
   &  &  & (0.239) & (0.242) & (0.241) & (0.248) & (0.247) \\ 
   &  &  &  &  &  &  &  \\ 
  Last U18 &  &  & 0.542 &  &  & 0.494 &  \\ 
   &  &  & (0.394) &  &  & (0.396) &  \\ 
   &  &  &  &  &  &  &  \\ 
  Last WJC &  &  & -0.131 &  &  & -0.245 &  \\ 
   &  &  & (0.336) &  &  & (0.344) &  \\ 
   &  &  &  &  &  &  &  \\ 
  Last WC &  &  & 0.115 &  &  & 0.225 &  \\ 
   &  &  & (0.265) &  &  & (0.273) &  \\ 
   &  &  &  &  &  &  &  \\ 
  Avg. Height &  &  &  & -0.195$^{**}$ &  & -0.123 & -0.142$^{*}$ \\ 
   &  &  &  & (0.082) &  & (0.083) & (0.082) \\ 
   &  &  &  &  &  &  &  \\ 
  Avg. Weight &  &  &  & 0.142$^{**}$ &  & 0.105$^{*}$ & 0.123$^{*}$ \\ 
   &  &  &  & (0.063) &  & (0.064) & (0.063) \\ 
   &  &  &  &  &  &  &  \\ 
  Avg. Age &  &  &  & 0.293 &  & 0.478 &  \\ 
   &  &  &  & (0.359) &  & (0.371) &  \\ 
   &  &  &  &  &  &  &  \\ 
  IIHF Exp. &  &  &  &  & 0.880$^{***}$ & 0.866$^{***}$ & 0.897$^{***}$ \\ 
   &  &  &  &  & (0.183) & (0.185) & (0.180) \\ 
   &  &  &  &  &  &  &  \\ 
  NHL Exp. &  &  &  &  & 0.070 & 0.084 &  \\ 
   &  &  &  &  & (0.059) & (0.060) &  \\ 
   &  &  &  &  &  &  &  \\ 
  Other Exp. &  &  &  &  & 0.009 & 0.007 &  \\ 
   &  &  &  &  & (0.008) & (0.008) &  \\ 
   &  &  &  &  &  &  &  \\ 
  Autoreg. Coef. &  & 0.850$^{***}$ & 0.834$^{***}$ & 0.788$^{***}$ & 0.852$^{***}$ & 0.800$^{***}$ & 0.796$^{***}$ \\ 
   &  & (0.121) & (0.152) & (0.242) & (0.105) & (0.190) & (0.177) \\ 
   &  &  &  &  &  &  &  \\ 
  Score Coef. &  & 0.070 & 0.059 & 0.060 & 0.079 & 0.060 & 0.059 \\ 
   &  & (0.048) & (0.062) & (0.058) & (0.051) & (0.069) & (0.056) \\ 
\midrule
Log-Lik. & -442.939 & -441.594 & -438.577 & -435.964 & -424.753 & \textbf{-420.814} & -424.078 \\ 
AIC & 913.877 & 915.189 & 917.153 & 911.927 & 889.506 & 893.628 & \textbf{888.157} \\ 
\bottomrule
\midrule
\multicolumn{8}{r}{$^{***}p < 0.01$; $^{**}p < 0.05$; $^{*}p < 0.1$} \\
\end{tabular}
\end{table}

\subsection{Forecasting Performance}
\label{sec:resFcst}

We evaluate the forecasting performance by re-estimating the models using the penalized maximum likelihood method with various values of the regularization parameter. We focus on the final specification of the predictor variables and compare it with the baseline static model. Rolling estimations with one-step-ahead forecasts are performed for the last 16 observations. We do not extend forecasts beyond the one-step-ahead horizon due to the unavailability of predictor variables for future tournaments. Even so, final rosters are announced just before the start of the tournaments, which affects the availability of the physical and experience variables. We compare the forecasts using several metrics, with the primary metric being the average log-likelihood of the one-step-ahead observation. To illustrate predictive ability, we assess three probabilities: the probability that the model correctly predicts the winner, the probability that the model correctly identifies the three medalists in any order, and the probability that the model correctly predicts which teams advance to the playoffs. Finally, we use the mean absolute error (MAE) and the root mean square error (RMSE) between the observed rankings and the modal rankings given by the model. We do not incorporate the mean ranking as it is computationally challenging to obtain. Therefore, the reported MAE and RMSE should be interpreted with caution.

The forecasting results are presented in Tables \ref{tab:fcstWC} and \ref{tab:fcstU20}. When penalization is employed, the final model consistently achieves a higher log-likelihood than the static model. However, without penalization, the static model outperforms the final model. Note that in rolling estimation, the partition condition of \cite{Hunter2004} does not always hold, making the use of penalization strongly advisable. Among the grid of values considered, the optimal regularization parameter $\lambda$ is found to be 0.01 for both WC and WJC, based on the average log-likelihood. The three probabilities studied reach their maximum values at lower $\lambda$ values; however, they remain relatively similar for $\lambda=0.01$. For WC, the final model with $\lambda=0.01$ correctly predicts the winner with a 17 percent probability, the three medalists with a 5 percent probability, and the eight teams advancing to the playoffs with a 3 percent probability. For WJC, the final model with $\lambda=0.01$ correctly predicts the winner with a 21 percent probability, the three medalists with a 7 percent probability, and the eight teams advancing to the playoffs with a 30 percent probability. Given that WC is played with 16 teams and WJC with 10 teams, the probabilities for advancing to the playoffs are naturally quite different. Regarding the evaluation of modal rankings, the lowest MAE and RMSE for WC are found with $\lambda=0.01$. For WJC, the lowest MAE is achieved with $\lambda=1$, while the lowest RMSE is achieved with $\lambda=0$, highlighting that the metrics based on the modal ranking may not be reliable. The MAE is 2.0 ranks for WC and 1.4 ranks for WJC when $\lambda=0.01$.

\newpage

\begin{table}
\caption{The average predicted log-likelihood, the probabilities of winning the tournament, earning a medal, and advancing to the playoff phase, as well as the mean absolute error and the root mean squared error for one-step-ahead forecasts of the World Championships. The static and final models are estimated using the penalized maximum likelihood method with various values of the regularization parameter. The best attained values are highlighted.}
\label{tab:fcstWC}
\centering
\resizebox{\textwidth}{!}{
\begin{tabular}{lcccccccccccccc}
\toprule
& \multicolumn{2}{c}{$\lambda=0$} & & \multicolumn{2}{c}{$\lambda=0.001$} & & \multicolumn{2}{c}{$\lambda=0.01$} & & \multicolumn{2}{c}{$\lambda=0.1$} & & \multicolumn{2}{c}{$\lambda=1$} \\  \cmidrule(l{3pt}r{3pt}){2-3} \cmidrule(l{3pt}r{3pt}){5-6} \cmidrule(l{3pt}r{3pt}){8-9} \cmidrule(l{3pt}r{3pt}){11-12} \cmidrule(l{3pt}r{3pt}){14-15}
Measure & Static & Final & & Static & Final & & Static & Final & & Static & Final & & Static & Final \\
\midrule
Log-Lik.    & -26.359 & -29.270 & & -23.385 & -23.413 & & -23.321 & \textbf{-22.783} & & -24.764 & -23.669 & & -28.778 & -27.793 \\ 
\\ 
P[Champion] & \textbf{0.172} & 0.170 & & 0.171 & 0.170 & & 0.166 & 0.166 & & 0.135 & 0.139 & & 0.079 & 0.083 \\ 
P[Medals]   & 0.032 & 0.050 & & 0.032 & \textbf{0.050} & & 0.028 & 0.045 & & 0.015 & 0.022 & & 0.003 & 0.004 \\  
P[Playoffs] & 0.028 & \textbf{0.033} & & 0.027 & 0.032 & & 0.022 & 0.027 & & 0.006 & 0.011 & & 0.000 & 0.001 \\ 
\\ 
MAE         & 2.173 & 2.030 & & 2.196 & 2.022 & & 2.212 & \textbf{1.999} & & 2.227 & 2.038 & & 2.594 & 2.023 \\
RMSE        & 2.766 & 2.657 & & 2.787 & 2.651 & & 2.812 & \textbf{2.627} & & 2.844 & 2.668 & & 3.314 & 2.647 \\ 
\bottomrule
\end{tabular}
}
\end{table}

\begin{table}
\caption{The average predicted log-likelihood, the probabilities of winning the tournament, earning a medal, and advancing to the playoff phase, as well as the mean absolute error and the root mean squared error for one-step-ahead forecasts of the World Junior Championships. The static and final models are estimated using the penalized maximum likelihood method with various values of the regularization parameter. The best attained values are highlighted.}
\label{tab:fcstU20}
\centering
\resizebox{\textwidth}{!}{
\begin{tabular}{lcccccccccccccc}
\toprule
& \multicolumn{2}{c}{$\lambda=0$} & & \multicolumn{2}{c}{$\lambda=0.001$} & & \multicolumn{2}{c}{$\lambda=0.01$} & & \multicolumn{2}{c}{$\lambda=0.1$} & & \multicolumn{2}{c}{$\lambda=1$} \\  \cmidrule(l{3pt}r{3pt}){2-3} \cmidrule(l{3pt}r{3pt}){5-6} \cmidrule(l{3pt}r{3pt}){8-9} \cmidrule(l{3pt}r{3pt}){11-12} \cmidrule(l{3pt}r{3pt}){14-15}
Measure & Static & Final & & Static & Final & & Static & Final & & Static & Final & & Static & Final \\
\midrule
Log-Lik.    & -12.457 & -13.380 & & -11.062 & -10.949 & & -11.004 & \textbf{-10.794} & & -11.745 & -11.072 & & -13.704 & -13.080 \\ 
\\ 
P[Champion] & 0.178 & \textbf{0.224} & & 0.178 & 0.219 & & 0.174 & 0.212 & & 0.153 & 0.180 & & 0.115 & 0.126 \\ 
P[Medals]   & 0.058 & 0.073 & & 0.058 & \textbf{0.075} & & 0.054 & 0.071 & & 0.036 & 0.049 & & 0.014 & 0.017 \\ 
P[Playoffs] & 0.283 & \textbf{0.332} & & 0.279 & 0.317 & & 0.251 & 0.303 & & 0.160 & 0.229 & & 0.067 & 0.091 \\ 
\\ 
MAE         & 1.606 & 1.337 & & 1.606 & 1.339 & & 1.606 & 1.353 & & 1.681 & 1.378 & & 1.822 & \textbf{1.314} \\ 
RMSE        & 2.081 & \textbf{1.763} & & 2.081 & 1.795 & & 2.081 & 1.805 & & 2.145 & 1.837 & & 2.391 & 1.796 \\ 
\bottomrule
\end{tabular}
}
\end{table}

\subsection{Unbounded Optimization}
\label{sec:resOpti}

\begin{figure}
\centering
\includegraphics[width=0.85\textwidth]{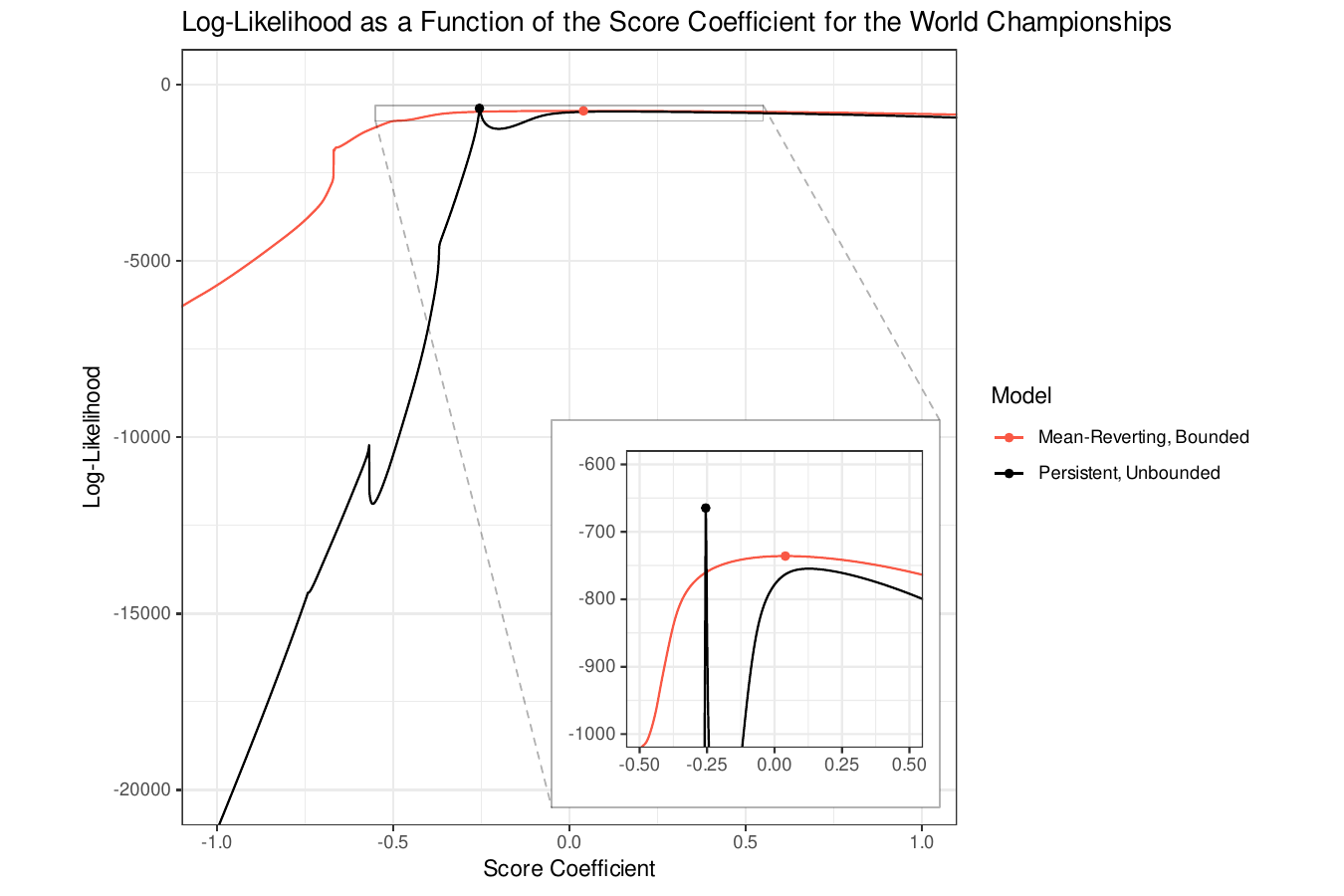}
\caption{The value of the log-likelihood for varying values of the score coefficient in the World Championships. Other coefficients are kept at their optimal values. The mean-reverting final model, with bounds $\varphi \in (0, 1)$ and $\alpha \geq 0$, and the persistent final model, with $\varphi=1$, are estimated using the maximum likelihood method.}
\label{fig:unboundedWC}
\end{figure}

\begin{figure}
\includegraphics[width=0.85\textwidth]{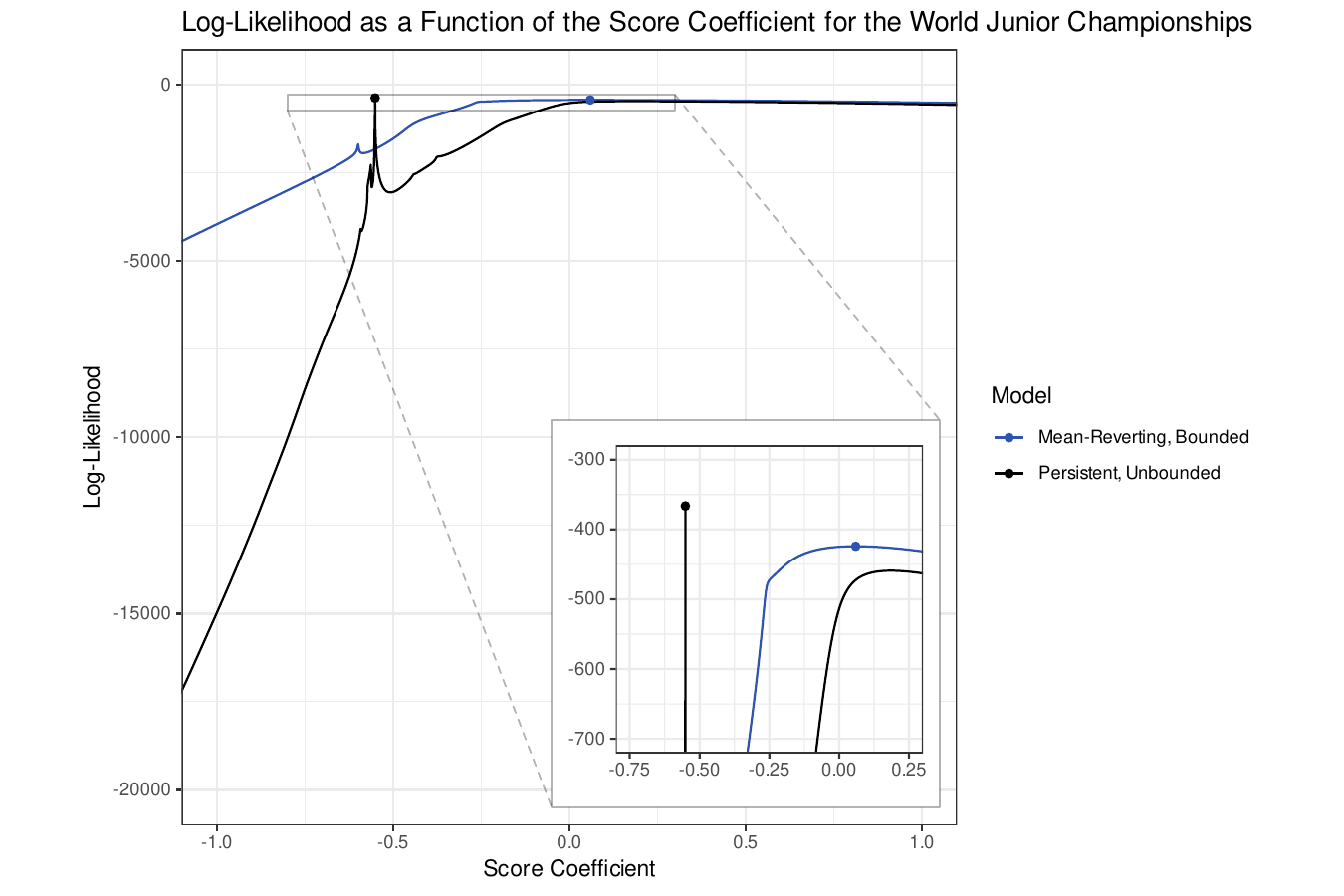}
\caption{The value of the log-likelihood for varying values of the score coefficient in the World Junior Championships. Other coefficients are kept at their optimal values. The mean-reverting final model, with bounds $\varphi \in (0, 1)$ and $\alpha \geq 0$, and the persistent final model, with $\varphi=1$, are estimated using the maximum likelihood method and its penalized version.}
\label{fig:unboundedU20}
\end{figure}

For the in-sample analysis in Section \ref{sec:resEst} and the out-of-sample analysis in Section \ref{sec:resFcst}, we employed bounds $\varphi \in [0, 1)$ and $\alpha \geq 0$. These bounds are quite natural, with $\varphi \geq 0$ and $\alpha \geq 0$ ensuring non-negative dependence on past strengths and tournament rankings, while $\varphi < 1$ is necessary for stationarity. However, when estimation is performed without these bounds, the optimal value of $\varphi$ tends to approach 1, and $\alpha$ becomes negative, leading to results that lack a straightforward interpretation. Figures \ref{fig:unboundedWC} and \ref{fig:unboundedU20} illustrate this issue for the final models of both WC and WJC. These figures show the log-likelihood as a function of the score coefficient, with the remaining coefficients kept at their optimal values. We investigate the mean-reverting model with bounds $\varphi \in [0, 1)$ and $\alpha \geq 0$, and the persistent model with $\varphi = 1$ and unrestricted $\alpha$. However, for the mean-reverting model, only the bound $\alpha \geq 0$ would suffice, as the optimal $\varphi$ then lies in $\varphi \in [0, 1)$. The persistent model resembles a random walk and is inherently non-stationary. In the positive region for both models, the log-likelihood functions appear to be concave. However, in the negative region, the functions exhibit irregular behavior, with several narrow peaks. In the case of unbounded estimation, one of these peaks becomes the global maximum. This irregular behavior poses significant challenges for numerical optimization algorithms, making it difficult to converge to this global maximum in the unbounded persistent model. Finding the global maximum may require multiple runs with different starting points. The persistent model with negative $\alpha$ is clearly overfitted, with even minor variations in $\alpha$, or any other coefficient, causing a significant drop in log-likelihood. Consequently, the estimated standard errors of all coefficients are close to zero, indicating an unreliable model. The forecasting performance of the persistent model with negative $\alpha$ is also notably poor. The average one-step-ahead log-likelihood is -25.723 for WC and -12.958 for WJC when $\lambda = 0.01$, which is much worse than that of the proposed mean-reverting final model and even the static model. The cause of this strange behavior remains unclear, but is remarkably widespread: it appears in our models for both WC and WJC, with and without penalization, and is also observed in the model of \cite{Holy2022f} on a WC data sample with a reduced time frame\footnote{However, \cite{Holy2022f} did not notice and report the optimality of the persistent model with negative $\alpha$. In their computations, they used the Nelder--Mead algortihm with starting values leading to a well-behaved mean-reverting model. They also estimated persistent model with $\varphi=1$, but again a positive $\alpha$ was found. Using different starting values, however, would lead to models with higher log-likelihood and negative $\alpha$ but also to several negative issues as discussed here. \cite{Holy2022f} therefore report results corresponding to the assumption of the $\alpha \geq 0$ bound.}. We recommend using bounds $\varphi \in [0, 1)$ and $\alpha \geq 0$ to ensure an interpretable and well-behaved model.

\section{Conclusion}
\label{sec:con}

The literature on sports statistics in ice hockey primarily focuses on the NHL, with considerably less attention given to international competitions. This paper aims to address this gap by analyzing the results of the World Championships and World Junior Championships. Specifically, we identify the factors driving team success and subsequently forecast performance. To achieve this, we employ a modified dynamic ranking model based on \cite{Holy2022f}.

The proposed methodology can be applied to any tournament and any team sport where final standings, i.e., complete rankings of all participating teams, are available. Future research could focus on challenging our assumption of the Plackett--Luce distribution and developing methods to include all teams in the sample without resorting to biased estimates through penalization. Both of these issues represent limitations of our study. Additionally, the behavior of the persistent model with a negative score coefficient warrants further investigation.

Our empirical analysis of international ice hockey reveals notable differences in the factors influencing performance in the WC compared to the WJC. Home advantage is significant only in the WJC, suggesting that younger players are more affected by external environments. This includes the size and shape of the rinks, which differ between Europe and North America. While the WJC is sometimes played on European rinks and sometimes on North American rinks, the WC is almost exclusively played on European rinks. In general, home advantage can result from various factors, such as familiarity with the arena, crowd support, reduced travel fatigue, and favorable scheduling. However, assessing the contribution of each factor can be challenging. A more in-depth investigation of the observed differences in home advantage between the WC and WJC could be an intriguing avenue for future research.

Success in the WJC positively influences subsequent performance in the WC, while success in the U18 is found to be insignificant for WJC performance. This is consistent with the view that the U18 serves as an evaluation process, showcasing young talent and emphasizing player development, whereas the WJC represents the peak of international junior competition, surrounded by prestige and fanfare. Our analysis reveals that the results in the WJC are even more indicative of future WC performance than the most recent results in the WC, highlighting the importance of this junior tournament.

In terms of physical attributes, shorter and heavier players on average had a significant advantage in the WJC, while in the WC, these traits were less impactful. In the early stages of development, such as in the WJC, body mass and stature may give players a competitive edge, particularly in a tournament where physicality often plays a more direct role due to the lower skill levels compared to elite professional leagues. However, as players progress to higher levels of competition, such as the WC, skill development, tactical awareness, and experience become more crucial, which may explain why the impact of physical traits diminishes. This could imply that coaches, scouts, and organizations prioritizing long-term player development should place less emphasis on height and weight and more on enhancing technical and mental skills that will allow players to succeed in higher-level competitions. Note that we examine the influence of these physical attributes regardless of player positions or game roles. Incorporating more detailed physical variables, such as those specific to forwards, defensemen, and goaltenders, could further improve the study.

Younger teams in the WC tend to perform better, provided they have sufficient experience. This suggests that younger players, often more agile and open to adopting innovative strategies, could exploit the fast-paced nature of the WC, where quick transitions and dynamic play are crucial. However, this must be complemented by experience from high-pressure competitions. From a practical standpoint, this finding underscores the importance of providing younger players with opportunities for international exposure and competitive play.

Another finding is that experience in past WCs significantly outweighs experience in the NHL. The WC and NHL are quite different tournaments: the WC is a more intense, short-term competition, while the NHL spans an entire season, culminating in the playoffs. Consequently, players in the WC often need to be more versatile, able to quickly adjust to different styles of play and high-pressure situations, while NHL players must maintain peak performance over a longer duration and under more physically demanding conditions. Our findings suggest that the skill sets required for success in the WC and NHL do indeed differ. However, this should be interpreted with caution, as our analysis offers only a limited comparison of NHL and WC experience. A key issue is that players from the best-performing NHL teams do not participate in the WC, as the NHL playoffs overlap with the WC. Our NHL experience variable does not account for this, which may introduce bias into the analysis. Future research could define and utilize more detailed experience variables, such as those reflecting players' success in the tournament rather than just their participation.

In forecasting, assuming stable strength levels over the long term and incorporating the proposed variables leads to the most accurate predictions. The potential use of more detailed variables, such as players' past achievements or characteristics aggregated at the level of player positions rather than the entire team, could improve the fit of the model; however, it could also lead to challenges related to overfitting. Especially considering that the information in the final standings is limited. It might be necessary to supplement the analysis of final standings with additional data, such as outcomes of individual matches or results from the Olympics. In our analysis, we utilize WC rankings from 1976 to 2024 and WJC rankings from 1977 to 2024. These are rather large time periods, which can raise concerns about temporal variation in the studied relations. For instance, the influence of home advantage might change over time, something our model is not designed to address. Future research could address this another potential issue.

As discussed above, our methodology could be extended in several ways. Nonetheless, the current study provides a unique analysis of international ice hockey, effectively balancing model simplicity, data structure, available information, and the range of predictor variables.

\section*{Funding}
\label{sec:fund}

The work on this paper was supported by the Czech Science Foundation under project 23-06139S and the personal and professional development support program of the Faculty of Informatics and Statistics, Prague University of Economics and Business.


\appendix

\section{Additional Tables and Figures}
\label{sec:tf}

\begin{figure}[H]
\centering
\includegraphics[width=\textwidth]{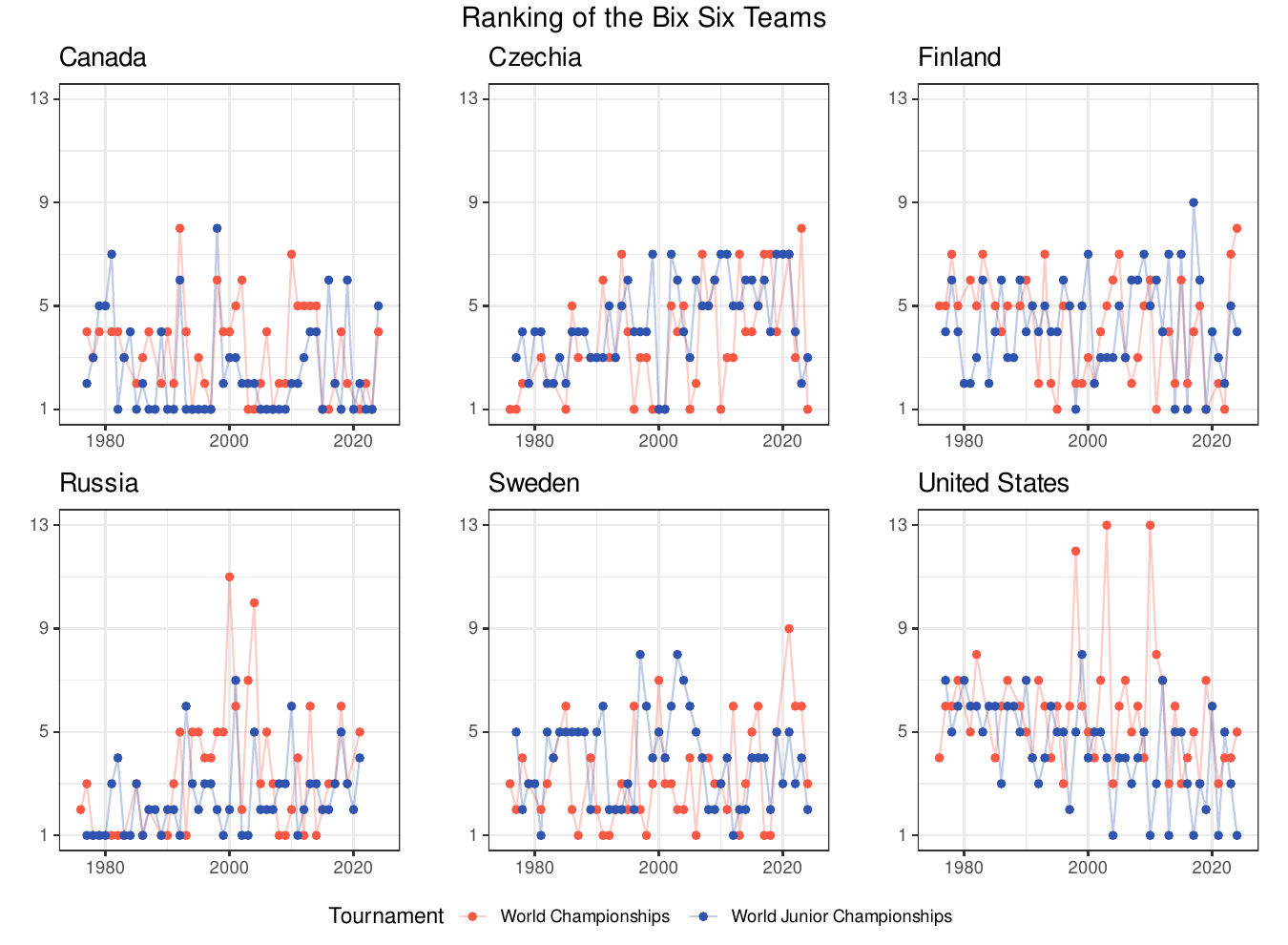}
\caption{The rankings of the Big Six teams in the World Championships (43rd--87th editions) and the World Junior Championships (1st--48th editions). The source of data is IIHF (\url{www.iihf.com}).}
\label{fig:rank}
\end{figure}

\begin{figure}
\centering
\includegraphics[width=\textwidth]{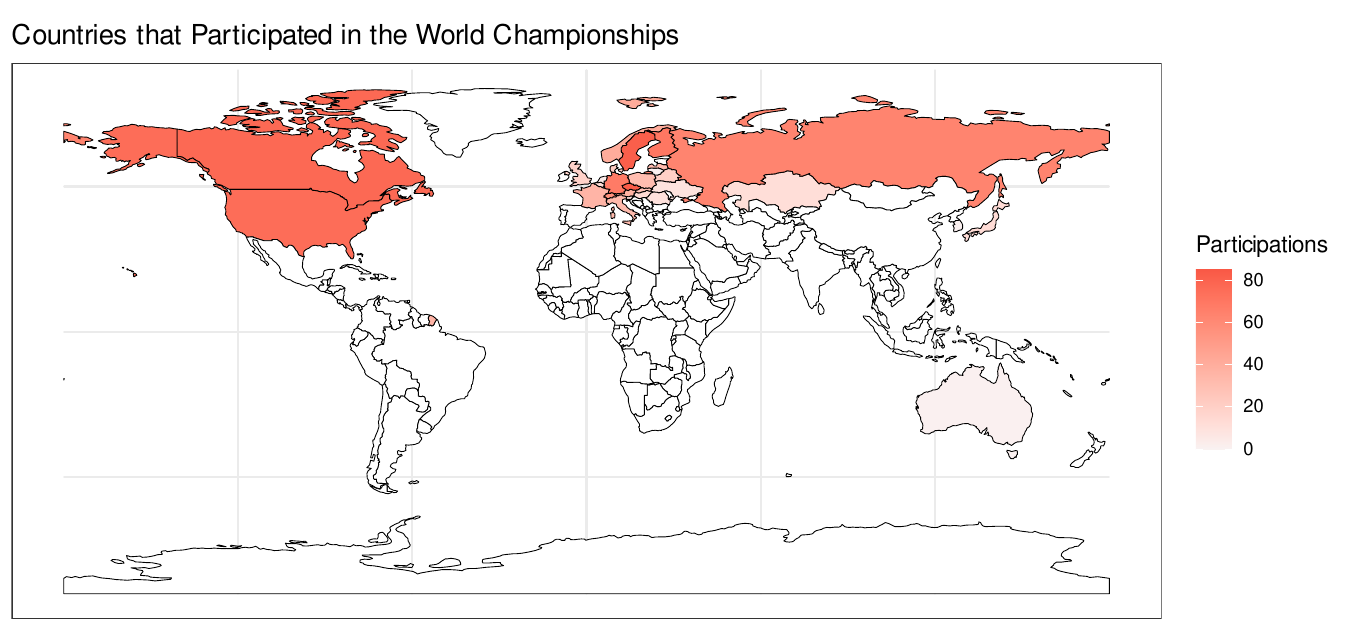}
\caption{The total number of times participated in the World Championships (1st--87th editions). The national teams are visualized according to their most recent borders. The former teams of East Germany and Yugoslavia are omitted. The source of data is IIHF (\url{www.iihf.com}).}
\label{fig:worldWC}
\end{figure}

\begin{figure}
\includegraphics[width=\textwidth]{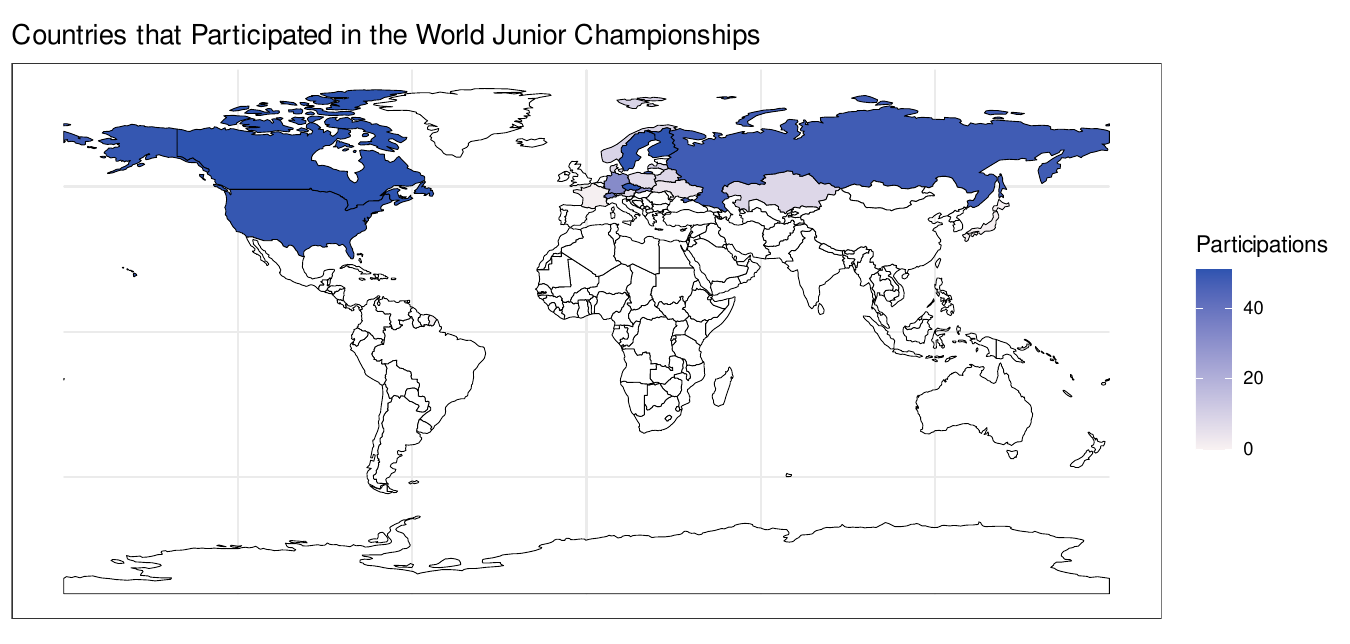}
\caption{The total number of times participated in the World Junior Championships (1st--48th editions). The national teams are visualized according to their most recent borders. The source of data is IIHF (\url{www.iihf.com}).}
\label{fig:worldU20}
\end{figure}

For more insight into WC and WJC, see Figures \ref{fig:rank}--\ref{fig:worldU20} and Tables \ref{tab:medals}--\ref{tab:effU20}.

\begin{table}[H]
\caption{The total number of medals by country in the World Championships (1st--87th editions), World Junior Championships (1st--48th editions), and World Under-18 Championships (1st--25th editions). The source of data is IIHF (\url{www.iihf.com}).}
\label{tab:medals}
\centering
\begin{tabular}{lccccccccccc}
\toprule
& \multicolumn{3}{c}{WC} & & \multicolumn{3}{c}{WJC} & & \multicolumn{3}{c}{U18} \\  \cmidrule(l{3pt}r{3pt}){2-4} \cmidrule(l{3pt}r{3pt}){6-8} \cmidrule(l{3pt}r{3pt}){10-12}
Team & \goldmedal & \silvermedal & \bronzemedal & & \goldmedal & \silvermedal & \bronzemedal & & \goldmedal & \silvermedal & \bronzemedal \\
\midrule
Canada & 28 & 16 & 9 & & 20 & 10 & 5 & & 5 & 1 & 4\\
Russia & 27 & 10 & 10 & & 13 & 13 & 11 & & 3 & 6 & 3 \\
Czechia & 13 & 13 & 22 & & 2 & 6 & 8 & & & 1 & 3\\
Sweden & 11 & 19 & 18 & & 2 & 12 & 7 & & 2 & 6 & 6 \\
Finland & 4 & 9 & 3 & & 5 & 5 & 7 & & 4 & 3 & 5 \\
United States & 2 & 9 & 9 & & 6 & 2 & 7 & & 11 & 6 & 3\\
\\ 
Great Britain & 1 & 2 & 2 \\
Slovakia & 1 & 2 & 1 & & & & 2 & & & 1 & 1 \\
Switzerland & & 4 & 8 & & & & 1 & & & 1 \\
Germany & & 3 & 2 \\
Austria & & & 2 \\
Latvia & & & 1 \\
\bottomrule
\end{tabular}
\end{table}

\newpage

\begin{table}
\caption{The five-number summary of the predictor variables for the World Championships.}
\label{tab:fiveWC}
\centering
\begin{tabular}{llrrrrr}
\toprule
Variable & Unit & Min & Q1 & Q2 & Q3 & Max \\ 
\midrule
Hosting & binary & 0.00 & 0.00 & 0.00 & 0.00 & 1.00 \\ 
\\ 
Last U18 & 1 / rank & 0.00 & 0.00 & 0.06 & 0.17 & 1.00 \\ 
Last WJC & 1 / rank & 0.00 & 0.05 & 0.14 & 0.25 & 1.00 \\ 
Last WC & 1 / rank & 0.00 & 0.08 & 0.12 & 0.25 & 1.00 \\ 
\\ 
Avg. Height & cm & 175.00 & 182.10 & 183.52 & 184.79 & 189.42 \\ 
Avg. Weight & kg & 75.65 & 86.50 & 88.13 & 90.16 & 94.96 \\ 
Avg. Age & years & 22.88 & 25.63 & 26.60 & 27.61 & 31.24 \\ 
\\ 
IIHF Exp. & games & 3.50 & 15.08 & 20.74 & 27.28 & 53.35 \\ 
NHL Exp. & games & 0.00 & 1.56 & 28.60 & 114.60 & 471.40 \\ 
Other Exp. & games & 0.00 & 30.73 & 89.05 & 208.83 & 438.24 \\
\bottomrule
\end{tabular}
\end{table}

\begin{table}
\caption{The five-number summary of the predictor variables for the World Junior Championships.}
\label{tab:fiveU20}
\centering
\begin{tabular}{llrrrrr}
\toprule
Variable & Unit & Min & Q1 & Q2 & Q3 & Max \\ 
\midrule
Hosting & binary & 0.00 & 0.00 & 0.00 & 0.00 & 1.00 \\ 
\\ 
Last U18 & 1 / rank & 0.00 & 0.00 & 0.00 & 0.20 & 1.00 \\ 
Last WJC & 1 / rank & 0.00 & 0.14 & 0.20 & 0.33 & 1.00 \\ 
Last WC & 1 / rank & 0.00 & 0.11 & 0.20 & 0.33 & 1.00 \\ 
\\ 
Avg. Height & cm & 176.17 & 182.17 & 183.87 & 185.21 & 187.78 \\ 
Avg. Weight & kg & 78.92 & 85.25 & 87.52 & 89.82 & 96.18 \\ 
Avg. Age & years & 18.40 & 19.12 & 19.26 & 19.40 & 19.75 \\ 
\\ 
IIHF Exp. & games & 0.00 & 0.00 & 0.00 & 0.40 & 2.10 \\ 
NHL Exp. & games & 0.00 & 0.00 & 0.00 & 0.22 & 12.00 \\ 
Other Exp. & games & 0.00 & 0.00 & 3.31 & 29.47 & 68.36 \\ 
\bottomrule
\end{tabular}
\end{table}

\begin{table}
\caption{The estimated fixed effects with the standard errors in parentheses for various models of the World Championships. The models are estimated by the maximum likelihood method.}
\label{tab:effWC}
\centering
\begin{tabular}{lccccccc}
\toprule
 & Static & Dynamic & Tourn. & Phys. & Exp. & Full & Final  \\
\midrule
  Austria & -0.942$^{***}$ & -0.933$^{***}$ & -0.819$^{***}$ & -1.007$^{***}$ & -1.104$^{***}$ & -0.776$^{**}$ & -0.820$^{***}$ \\ 
   & (0.280) & (0.342) & (0.309) & (0.348) & (0.325) & (0.316) & (0.299) \\ 
   &  &  &  &  &  &  &  \\ 
  Belarus & -0.164 & -0.180 & -0.078 & -0.302 & -0.341 & -0.240 & -0.179 \\ 
   & (0.283) & (0.366) & (0.320) & (0.383) & (0.332) & (0.323) & (0.309) \\ 
   &  &  &  &  &  &  &  \\ 
  Canada & 3.099$^{***}$ & 3.080$^{***}$ & 2.527$^{***}$ & 2.807$^{***}$ & 3.499$^{***}$ & 2.225$^{***}$ & 2.268$^{***}$ \\ 
   & (0.232) & (0.292) & (0.297) & (0.336) & (0.453) & (0.505) & (0.482) \\ 
   &  &  &  &  &  &  &  \\ 
  Czechia & 2.887$^{***}$ & 2.940$^{***}$ & 2.654$^{***}$ & 2.725$^{***}$ & 2.868$^{***}$ & 2.207$^{***}$ & 2.380$^{***}$ \\ 
   & (0.226) & (0.290) & (0.266) & (0.314) & (0.289) & (0.307) & (0.262) \\ 
   &  &  &  &  &  &  &  \\ 
  Denmark & -0.073 & -0.094 & 0.035 & 0.005 & -0.368 & -0.214 & -0.334 \\ 
   & (0.258) & (0.358) & (0.303) & (0.372) & (0.321) & (0.306) & (0.295) \\ 
   &  &  &  &  &  &  &  \\ 
  East Germany & -1.713 & -1.751 & -1.428 & -1.555 & -1.296 & -1.144 & -1.106 \\ 
   & (1.362) & (1.348) & (1.272) & (1.323) & (1.272) & (1.214) & (1.235) \\ 
   &  &  &  &  &  &  &  \\ 
  Finland & 2.425$^{***}$ & 2.368$^{***}$ & 2.138$^{***}$ & 2.249$^{***}$ & 2.303$^{***}$ & 1.605$^{***}$ & 1.852$^{***}$ \\ 
   & (0.229) & (0.286) & (0.262) & (0.295) & (0.344) & (0.364) & (0.254) \\ 
   &  &  &  &  &  &  &  \\ 
  France & -0.744$^{***}$ & -0.796$^{**}$ & -0.607$^{**}$ & -0.642$^{*}$ & -1.116$^{***}$ & -0.618$^{**}$ & -0.756$^{***}$ \\ 
   & (0.258) & (0.321) & (0.289) & (0.333) & (0.311) & (0.311) & (0.293) \\ 
   &  &  &  &  &  &  &  \\ 
  Germany & 0.619$^{***}$ & 0.605$^{**}$ & 0.645$^{***}$ & 0.612$^{**}$ & 0.480$^{*}$ & 0.304 & 0.398$^{*}$ \\ 
   & (0.218) & (0.283) & (0.246) & (0.290) & (0.282) & (0.276) & (0.239) \\ 
   &  &  &  &  &  &  &  \\ 
  Great Britain & -1.938$^{***}$ & -2.005$^{***}$ & -1.794$^{***}$ & -1.830$^{***}$ & -1.849$^{***}$ & -0.828 & -1.061 \\ 
   & (0.640) & (0.692) & (0.666) & (0.701) & (0.714) & (0.759) & (0.716) \\ 
   &  &  &  &  &  &  &  \\ 
  Hungary & -2.150$^{***}$ & -2.126$^{***}$ & -1.966$^{**}$ & -2.108$^{**}$ & -1.759$^{**}$ & -1.186 & -1.384 \\ 
   & (0.808) & (0.820) & (0.815) & (0.829) & (0.830) & (0.881) & (0.860) \\ 
   &  &  &  &  &  &  &  \\ 
  Italy & -0.996$^{***}$ & -0.907$^{***}$ & -0.792$^{**}$ & -0.763$^{**}$ & -1.148$^{***}$ & -0.677$^{*}$ & -0.764$^{**}$ \\ 
   & (0.291) & (0.350) & (0.320) & (0.373) & (0.342) & (0.353) & (0.321) \\
\bottomrule
\midrule
\multicolumn{8}{r}{$^{***}p < 0.01$; $^{**}p < 0.05$; $^{*}p < 0.1$} \\
\end{tabular}
\end{table}

\begin{table}
\caption*{Table \ref*{tab:effWC} (Continued): The estimated fixed effects with the standard errors in parentheses for various models of the World Championships. The models are estimated by the maximum likelihood method.}
\centering
\begin{tabular}{lccccccc}
\toprule
 & Static & Dynamic & Tourn. & Phys. & Exp. & Full & Final  \\
\midrule
Japan & -2.930$^{***}$ & -2.811$^{***}$ & -2.704$^{***}$ & -2.166$^{**}$ & -2.812$^{***}$ & -2.221$^{***}$ & -2.469$^{***}$ \\ 
   & (0.730) & (0.806) & (0.763) & (0.879) & (0.804) & (0.842) & (0.767) \\ 
   &  &  &  &  &  &  &  \\ 
  Kazakhstan & -1.740$^{***}$ & -1.761$^{***}$ & -1.611$^{***}$ & -1.583$^{***}$ & -1.568$^{***}$ & -1.000$^{**}$ & -1.130$^{**}$ \\ 
   & (0.429) & (0.464) & (0.447) & (0.471) & (0.454) & (0.457) & (0.448) \\ 
   &  &  &  &  &  &  &  \\ 
  Latvia & 0.371 & 0.417 & 0.494$^{*}$ & 0.279 & -0.005 & 0.057 & 0.128 \\ 
   & (0.236) & (0.329) & (0.279) & (0.346) & (0.307) & (0.292) & (0.282) \\ 
   &  &  &  &  &  &  &  \\ 
  Norway & -0.227 & -0.244 & -0.104 & -0.312 & -0.588$^{**}$ & -0.564$^{*}$ & -0.553$^{*}$ \\ 
   & (0.241) & (0.310) & (0.274) & (0.321) & (0.298) & (0.288) & (0.283) \\ 
   &  &  &  &  &  &  &  \\ 
  Poland & -2.465$^{***}$ & -2.405$^{***}$ & -2.329$^{***}$ & -2.329$^{***}$ & -2.169$^{***}$ & -1.831$^{**}$ & -1.838$^{**}$ \\ 
   & (0.766) & (0.773) & (0.778) & (0.778) & (0.779) & (0.795) & (0.778) \\ 
   &  &  &  &  &  &  &  \\ 
  Russia & 2.962$^{***}$ & 3.189$^{***}$ & 2.452$^{***}$ & 2.818$^{***}$ & 2.877$^{***}$ & 1.963$^{***}$ & 2.109$^{***}$ \\ 
   & (0.235) & (0.307) & (0.327) & (0.349) & (0.280) & (0.337) & (0.307) \\ 
   &  &  &  &  &  &  &  \\ 
  Slovakia & 0.994$^{***}$ & 1.011$^{***}$ & 0.999$^{***}$ & 0.899$^{***}$ & 0.996$^{***}$ & 0.861$^{***}$ & 0.880$^{***}$ \\ 
   & (0.243) & (0.329) & (0.280) & (0.347) & (0.290) & (0.280) & (0.261) \\ 
   &  &  &  &  &  &  &  \\ 
  Slovenia & -2.317$^{***}$ & -2.323$^{***}$ & -2.158$^{***}$ & -2.215$^{***}$ & -2.240$^{***}$ & -1.801$^{***}$ & -1.948$^{***}$ \\ 
   & (0.500) & (0.541) & (0.521) & (0.548) & (0.523) & (0.523) & (0.509) \\ 
   &  &  &  &  &  &  &  \\ 
  Sweden & 3.032$^{***}$ & 2.971$^{***}$ & 2.633$^{***}$ & 2.983$^{***}$ & 3.209$^{***}$ & 2.515$^{***}$ & 2.622$^{***}$ \\ 
   & (0.229) & (0.288) & (0.275) & (0.304) & (0.314) & (0.348) & (0.276) \\ 
   &  &  &  &  &  &  &  \\ 
  Switzerland & 1.028$^{***}$ & 0.936$^{***}$ & 0.978$^{***}$ & 0.895$^{***}$ & 0.877$^{**}$ & 0.419 & 0.675$^{**}$ \\ 
   & (0.244) & (0.313) & (0.275) & (0.324) & (0.364) & (0.372) & (0.272) \\ 
   &  &  &  &  &  &  &  \\ 
  Ukraine & -0.701$^{*}$ & -0.786 & -0.598 & -0.838 & -0.708 & -0.257 & -0.248 \\ 
   & (0.408) & (0.508) & (0.457) & (0.529) & (0.473) & (0.463) & (0.440) \\ 
   &  &  &  &  &  &  &  \\ 
  United States & 1.683$^{***}$ & 1.604$^{***}$ & 1.432$^{***}$ & 1.378$^{***}$ & 1.963$^{***}$ & 1.203$^{***}$ & 1.279$^{***}$ \\ 
   & (0.223) & (0.285) & (0.281) & (0.316) & (0.317) & (0.358) & (0.329) \\
\bottomrule
\midrule
\multicolumn{8}{r}{$^{***}p < 0.01$; $^{**}p < 0.05$; $^{*}p < 0.1$} \\
\end{tabular}
\end{table}

\begin{table}
\caption{The estimated fixed effects with the standard errors in parentheses for various models of the World Junior Championships. The models are estimated by the maximum likelihood method.}
\label{tab:effU20}
\centering
\begin{tabular}{lccccccc}
\toprule
 & Static & Dynamic & Tourn. & Phys. & Exp. & Full & Final  \\
\midrule
  Belarus & -3.159$^{***}$ & -3.133$^{***}$ & -3.054$^{***}$ & -2.932$^{***}$ & -3.144$^{***}$ & -2.994$^{***}$ & -3.193$^{***}$ \\ 
   & (0.975) & (0.988) & (0.985) & (0.992) & (1.005) & (1.010) & (0.997) \\ 
   &  &  &  &  &  &  &  \\ 
  Canada & 2.706$^{***}$ & 2.663$^{***}$ & 2.704$^{***}$ & 2.593$^{***}$ & 2.806$^{***}$ & 2.625$^{***}$ & 2.797$^{***}$ \\ 
   & (0.249) & (0.285) & (0.317) & (0.314) & (0.345) & (0.398) & (0.318) \\ 
   &  &  &  &  &  &  &  \\ 
  Czechia & 1.741$^{***}$ & 1.808$^{***}$ & 1.745$^{***}$ & 1.635$^{***}$ & 1.867$^{***}$ & 1.694$^{***}$ & 1.791$^{***}$ \\ 
   & (0.230) & (0.271) & (0.271) & (0.278) & (0.291) & (0.301) & (0.283) \\ 
   &  &  &  &  &  &  &  \\ 
  Denmark & -1.399$^{**}$ & -1.395$^{**}$ & -1.394$^{**}$ & -1.313$^{**}$ & -1.375$^{**}$ & -1.185$^{*}$ & -1.394$^{**}$ \\ 
   & (0.599) & (0.627) & (0.641) & (0.633) & (0.644) & (0.650) & (0.631) \\ 
   &  &  &  &  &  &  &  \\ 
  Finland & 1.656$^{***}$ & 1.644$^{***}$ & 1.532$^{***}$ & 1.518$^{***}$ & 1.307$^{***}$ & 1.226$^{***}$ & 1.530$^{***}$ \\ 
   & (0.231) & (0.267) & (0.273) & (0.260) & (0.372) & (0.366) & (0.261) \\ 
   &  &  &  &  &  &  &  \\ 
  Germany & -0.771$^{***}$ & -0.777$^{***}$ & -0.763$^{***}$ & -0.638$^{**}$ & -0.804$^{**}$ & -0.692$^{**}$ & -0.651$^{**}$ \\ 
   & (0.274) & (0.300) & (0.295) & (0.296) & (0.312) & (0.309) & (0.301) \\ 
   &  &  &  &  &  &  &  \\ 
  Kazakhstan & -0.981$^{*}$ & -0.973$^{*}$ & -0.858 & -0.447 & -0.751 & -0.228 & -0.381 \\ 
   & (0.549) & (0.569) & (0.574) & (0.643) & (0.582) & (0.652) & (0.645) \\ 
   &  &  &  &  &  &  &  \\ 
  Latvia & -1.346$^{***}$ & -1.338$^{**}$ & -1.422$^{**}$ & -1.331$^{**}$ & -1.612$^{***}$ & -1.419$^{**}$ & -1.645$^{***}$ \\ 
   & (0.513) & (0.538) & (0.553) & (0.552) & (0.574) & (0.581) & (0.568) \\ 
   &  &  &  &  &  &  &  \\ 
  Norway & -2.440$^{***}$ & -2.443$^{***}$ & -2.329$^{***}$ & -2.333$^{***}$ & -2.604$^{***}$ & -2.541$^{***}$ & -2.730$^{***}$ \\ 
   & (0.726) & (0.731) & (0.732) & (0.734) & (0.755) & (0.768) & (0.770) \\ 
   &  &  &  &  &  &  &  \\ 
  Poland & -2.249$^{**}$ & -2.298$^{**}$ & -2.305$^{**}$ & -2.529$^{**}$ & -2.085$^{**}$ & -2.323$^{**}$ & -2.274$^{**}$ \\ 
   & (1.009) & (1.021) & (1.019) & (1.033) & (1.018) & (1.029) & (1.028) \\ 
   &  &  &  &  &  &  &  \\ 
  Russia & 2.847$^{***}$ & 2.856$^{***}$ & 2.811$^{***}$ & 2.578$^{***}$ & 2.998$^{***}$ & 2.714$^{***}$ & 2.720$^{***}$ \\ 
   & (0.252) & (0.288) & (0.320) & (0.315) & (0.299) & (0.361) & (0.319) \\ 
   &  &  &  &  &  &  &  \\ 
  Slovakia & 0.215 & 0.215 & 0.236 & 0.283 & 0.217 & 0.260 & 0.154 \\ 
   & (0.264) & (0.310) & (0.304) & (0.305) & (0.330) & (0.325) & (0.310) \\ 
   &  &  &  &  &  &  &  \\ 
  Sweden & 1.908$^{***}$ & 1.953$^{***}$ & 1.893$^{***}$ & 1.835$^{***}$ & 1.813$^{***}$ & 1.656$^{***}$ & 1.967$^{***}$ \\ 
   & (0.235) & (0.273) & (0.280) & (0.274) & (0.344) & (0.344) & (0.269) \\ 
   &  &  &  &  &  &  &  \\ 
  Switzerland & -0.296 & -0.356 & -0.311 & -0.395 & -0.338 & -0.307 & -0.230 \\ 
   & (0.262) & (0.295) & (0.297) & (0.291) & (0.329) & (0.332) & (0.296) \\ 
   &  &  &  &  &  &  &  \\ 
  United States & 1.567$^{***}$ & 1.574$^{***}$ & 1.515$^{***}$ & 1.477$^{***}$ & 1.704$^{***}$ & 1.513$^{***}$ & 1.539$^{***}$ \\ 
   & (0.236) & (0.270) & (0.278) & (0.274) & (0.287) & (0.297) & (0.277) \\
\bottomrule
\midrule
\multicolumn{8}{r}{$^{***}p < 0.01$; $^{**}p < 0.05$; $^{*}p < 0.1$} \\
\end{tabular}
\end{table}

\end{document}